\documentclass[aps, prb,twocolumn,superscriptaddress]{revtex4}

\usepackage{color}
\usepackage{hyperref}
\usepackage{graphicx}
\usepackage{natbib}

\usepackage{amsmath}

\newcommand{\e}[1]{\mathrm{e}^{#1}}
\newcommand{\mr}[1]{\mathrm{#1}}
\newcommand{\mb}[1]{\mathbf{#1}}
\newcommand{\up}[0]{\uparrow}
\newcommand{\dn}[0]{\downarrow}

\newcommand{\EqLabel}[1]{\label{#1}}

\begin{document}

\title{The magnon-mediated attraction between two holes doped in a CuO$_2$ layer} 

\author{Mirko M. M\"oller} 
\affiliation{Department of Physics and Astronomy, University of
British Columbia, Vancouver B.C. V6T 1Z1, Canada} 
\affiliation{Stewart Blusson Quantum Matter Institute, University of
British Columbia, Vancouver B.C. V6T 1Z4, Canada}

\author{Clemens P.J. Adolphs}
\affiliation{1QBit, Vancouver B.C.  V6C 2B5, Canada}

\author{Mona Berciu}
\affiliation{Department of Physics and Astronomy, University of
British Columbia, Vancouver B.C. V6T 1Z1, Canada}
\affiliation{Stewart Blusson Quantum Matter Institute, University of
British Columbia, Vancouver B.C. V6T 1Z4, Canada}

\begin{abstract} 
  Using a realistic multi-band model for two holes doped into a
  CuO$_2$ layer, we devise a method to turn off the magnon-mediated
  interaction between the holes. This allows us to verify that this
  interaction is attractive, and therefore could indeed be (part of)
  the superconducting glue. We derive its analytical expression and
  show that it consists of a novel kind of pair-hopping+spin-exchange
  terms. Its coupling constant is fitted from the ground state energy
  obtained with variational exact diagonalization, and it faithfully
  reproduces the effect of the magnon-mediated attraction in the
  entire Brillouin zone. For realistic parameter values, this
  effective interaction is borderline strong enough to bind the holes
  into preformed pairs.
\end{abstract}

\maketitle

\section{Introduction}

Despite sustained efforts, more than thirty years after the discovery
of high-temperature superconductivity in cuprates,\cite{BedMu} the
nature of the glue that binds its Cooper pairs is still unclear.
\cite{Anderson2007} This binding is not through the
phonon-mediated Bardeen-Cooper-Schrieffer (BCS) mechanism responsible
for low-T$_c$, conventional superconductivity,\cite{BCS} although phonons have
been proposed as the glue for bipolaron superconductivity.
\cite{AMbip, Alexandrov2009,Kresin} The current leading contender
appears to be a magnon glue \cite{Mospfl,rev1,rev2,Mont,Valkov} due
to the proximity of antiferromagnetism in the phase diagram of these strongly
correlated materials, and also because of the existence of several
other non-conventional superconductors with an adjacent
magnetically-ordered phase.\cite{pnic,heavyfermions} Other,
more exotic  proposed glues include loop currents,\cite{Varma} orbital
relaxation \cite{Hirsch} or hidden fermions.\cite{Imada} A mix of
several glues is certainly also possible. \cite{Cataudella1,Conte}

Part of the reason for the absence of a definitive theoretical answer
is the fact that most such work is based on effective models or 
on phenomenological considerations, with parameters extracted from
fits of various experimental measurements. Such approaches are {\em a
  priori} guaranteed to reproduce some experimental aspects, but it is
not clear if the values of the fitted parameters are reasonable, nor how they
should depend on the microscopic structure or on external
parameters such as pressure, doping, {\em etc}. Such theories are hard
to falsify.

What is needed, instead, is to extract the form and the strength of
the effective attraction mediated by various glues, starting from
well-established microscopic models. In a second stage, these
effective interactions should then be investigated to see if they can
explain the high-T$_C$ superconductivity (and hopefully many other
aspects of the complex cuprate phenomenology) on their own, or if
combinations of several such terms are necessary. Clearly, this would
make the process of validating or falsifying various mechanisms more
straightforward. The problem, however, is that extracting these
effective attractions from microscopic models is very difficult, for
two reasons: (i) perturbative methods are unsuitable whether one
believes the cuprates to be strongly-correlated electron systems,
and/or to have the strong electron-phonon coupling that could enable a
high T$_C$ value. Moreover, one cannot appeal (only) to numerical methods to
obtain the needed analytical expressions for these effective
attractions. Instead, accurate (semi)analytical formalisms are needed,
and those are hard to come by; (ii) more fundamental is a problem
stemming from the indistinguishability of electrons. The effective
interactions arise from processes where one particle emits a boson,
which is then absorbed by another particle. If one could turn off this
process ``by hand'' and thus compare results where this
exchange is allowed {\em vs.} forbidden, one could infer the form and
magnitude of this effective boson-mediated interaction from its
effects on the many-body spectrum and wavefunctions. The problem is
that for indistinguishable particles it is impossible to tell which is 
``particle 1'' and which is ``particle 2'', in other words to distinguish
whether a boson has been exchanged or whether it has been
re-absorbed by the same particle that emitted it (thereby contributing to renormalizing it
into a quasiparticle, instead of to the boson-mediated interaction).

In this work we propose an elegant solution for these challenges that
allows us to verify that magnon-exchange indeed mediates an effective
attraction between two holes doped into a cuprate layer. Moreover, we
find the analytical expression of this effective attraction. Our
expression describes processes that are conceptually simple, namely pair-hopping+exchange terms where both holes hop
while also exchanging their spins. To the best of our knowledge, this
type of effective interaction has not been considered before in this
context. We extract its energy scale by fitting the ground-state
energy, specifically we ask that the ground-state energy of the system
with magnon-exchange allowed is reproduced by that of the system where
the magnon-exchange is turned off, but this additional effective
attraction is added instead. We then show that our effective
interaction reproduces well the effects of the magnon-exchange
throughout the Brillouin zone, thus validating its expression and
magnitude.

The paper is organized as follows. Section II presents the model and
describes its two-hole spectrum. Section III discusses how we prove
the existence of a magnon-mediated attraction between the holes, and
how we quantify its form and magnitude. Section IV analyzes the role
of the background spin fluctuations, while Section V speculates on the
possible existence of pre-formed pairs. Finally, Section VI contains a
short summary and conclusions. Various technical details are relegated
to the three Appendixes.

\section{The model and its two-hole spectrum}

It is well-known that in the doped cuprates, the doped holes reside in
the O $2p$ band, thus a reasonable starting point for an
accurate description is the three-band Emery model
\cite{Emery,mod4} which includes both the Cu $3d_{x^2-y^2}$ orbitals
that host the strongly-correlated holes responsible for the long-range
antiferromagnetic (AFM) order in the parent compounds, but also the
ligand O $2p$ orbitals hosting the additional doped holes responsible
for superconductivity.

\begin{figure}[t]
  \includegraphics[width=\columnwidth]{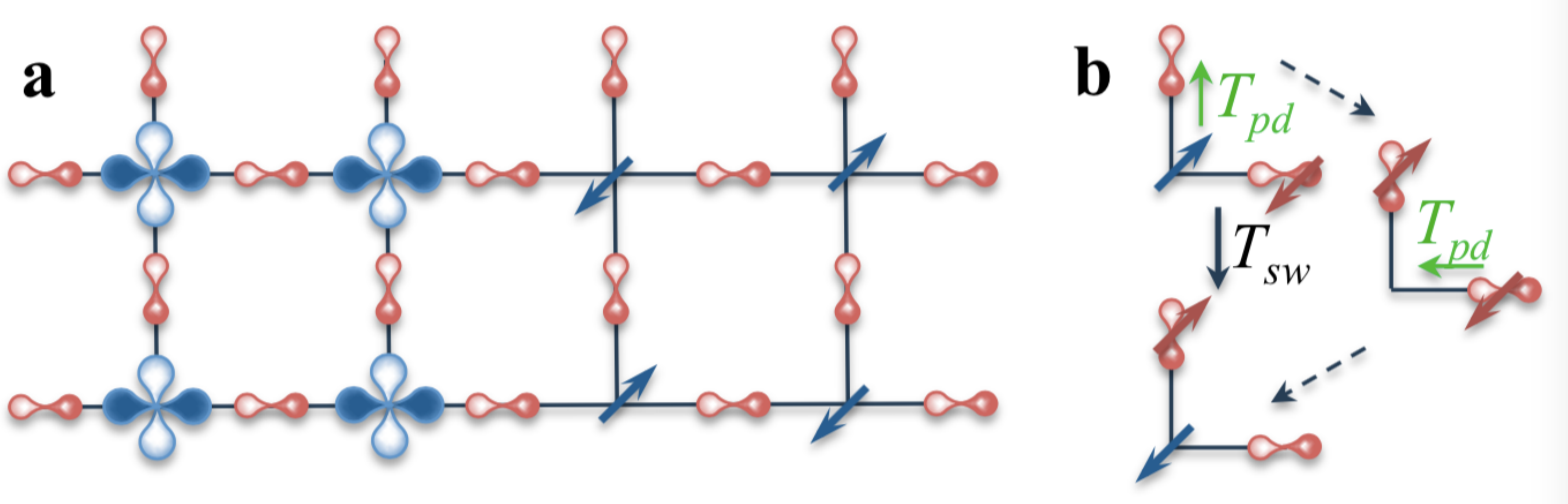}
\caption{(a) Sketch of three-band model which includes the Cu
  $3d_{x^2-y^2}$ and the O ligand $2p_{x/y}$ orbitals (left half). In the strongly-correlated limit, there are spin degrees of freedom at Cu sites while the doped holes move on the O sublattice, as sketched in the right half; (b) A $T_{sw}$  process which results in effective hopping of the  hole while its spin is swapped with that of the neighbour Cu.}
\label{fig1}
\end{figure}

We study the $U_{dd}\rightarrow \infty$ limit of the three-band Emery
model. This is justified physically because $U_{dd}$ is by far the
largest energy scale, and is necessary computationally to make the
Hilbert space manageable. The $U_{dd}\rightarrow \infty$ limit implies
that single-hole occupancy is enforced for the Cu $3d_{x^2-y^2}$
orbitals, so there are spins-${1\over 2}$ at these sites. Additional
(doped) holes occupy states in the O-$2p$ band derived from the ligand
$2p$ orbitals, as sketched on the right-hand side of Fig.
\ref{fig1}(a).

We believe this to be a more suitable starting point than the more
studied one-band $t-J$ and Hubbard models because the one-band models
make the additional assumption that the doped holes are locked into
Zhang-Rice singlets (ZRS) \cite{ZR,SE} and perform a further
projection onto those states. Even if the ZRS provides a good
description of a single quasiparticle ({\em qp}), it would not
necessarily follow that modeling the many-hole system in terms of ZRS
is valid. Nearby holes may modify the magnetic background and exchange
magnons in a way not allowed if each hole is locked in a
ZRS.\cite{MirkoPRL} Our approach has fewer constraints as it does not
impose the formation of ZRS, although it allows it to occur if it
turns out to be the most energetically favorable option. By being more
general, our model allows for, and tests, more possible scenarios.

Moreover, in previous work \cite{HadiNP, HadiJCP} we showed a
qualitative difference between the quasiparticle ({\em qp}) of our
model and that of the optimized $t$-$t'$-$t''$-$J$ model: while both models
predict a {\em qp} in agreement with that measured experimentally, the
dispersion in the one-band models is significantly impacted by
background spin fluctuations, unlike that of our model. The physical
origin of this difference is discussed below in detail. Here we note
that its existence suggests that these models do not describe the same
physics even in the single {\em qp} sector, so there is no reason to
expect them to describe the same magnon-mediated exchange in the
two-hole sector.

To obtain the many-hole Hamiltonian, we start from the Emery model and take its $U_{dd}\rightarrow \infty$ limit by straightforward generalization of the method used for
one hole in Ref. \onlinecite{Bayo}. The resulting  Hamiltonian is:
\begin{equation}
\EqLabel{e1} {\cal H} = T_{pp} + U_{pp}+ T_{sw} + H_{J_{pd}}
+H_{J_{dd}}.
\end{equation}
Briefly, $T_{pp}$ includes first and second nearest neighbour (nn)
hopping of the doped holes between ligand O $2p$ orbitals, while
$U_{pp}$ is the corresponding on-site repulsion. $T_{sw}$ describes
effective hopping of doped holes mediated by the Cu spin, whereby the
Cu hole hops onto a neighbour O followed by the doped hole filling the
Cu orbital, as sketched in Fig. \ref{fig1}(b); this leads to a swap of the spins
of the hole and the Cu. $H_{J_{pd}}$ is the AFM exchange between the
spins of the doped holes and those of their neighbouring Cu. Finally,
$H_{J_{dd}}$ is the nn AFM superexchange between adjacent Cu spins,
apart from bonds occupied by holes. Setting $J_{dd}\approx 150$meV as
the energy unit, we find $t_{pp} = 4.13, t_{pp}'=2.40, U_{dd}=25.40,
t_{sw}=2.98$ and $J_{pd}=2.83$, respectively.
\cite{Bayo,HadiNP,HadiJCP} Note that the value of $t_{sw}$ is changed
if a second hole is on either O involved, because $U_{pp}$ shifts the
energy of the intermediary states. This is taken into account in our calculations, although we found it to have
essentially no consequences. The detailed description of all these
terms and several other relevant technical details are given in Appendix
\ref{app1}.

In the undoped system, only the AFM superexchange $H_{J_{dd}}$ acts
between neighbor Cu spins. This Heisenberg-type exchange results in a
very complicated undoped ground-state (GS), which has strong
short-range AFM fluctuations but no long-range order. This is unlike
the real materials, which acquire long-range AFM order due to coupling
to neighboring layers.

To make progress, we begin by simplifying $H_{J_{dd}}$ to an Ising
form, so that the undoped ground-state is a Ne\'el state without spin
fluctuations. At first sight, it may seem counterintuitive that this is a reasonable
approximation (even though it produces a GS with long-range order,
much more similar to that of the actual material than is the GS of the
Heisenberg model). In fact, this turns out to be an excellent
approximation for this model, so far as the behavior of doped holes is
concerned, at least in the extremely underdoped limit we study here.
Indeed, as shown in Refs. \onlinecite{HadiNP, HadiJCP} for a single
doped hole, this approximation is justified because $J_{dd}$ is
significantly smaller than all other energy scales. Physically, this
means that the time-scale over which the background spin fluctuations
occur is significantly longer than that over which the holes move
around and modify their local magnetic environment and exchange
magnons through the spin-offdiagonal parts of $T_{sw}$ and
$H_{J_{pd}}$. Because spin fluctuations are so slow, their influence
on these fast processes involving the holes is minor. Below, we verify
explicitly that this holds true for the magnon-mediated interaction
between the holes, by allowing background spin-fluctuations to occur
in the vicinity of the holes. As discussed later, we find that their
presence changes the magnitude of the magnon-mediated interactions by only
a few percent, so indeed they are negligible. As 
mentioned, this is in sharp contrast to what happens in one-hole
models, where the spin-fluctuations occur on time-scales comparable to
those relevant for processes involving the doped holes, so 
they significantly influence their dispersion (and presumably the effective
interations, too).\cite{HadiJCP}

In the absence of background spin fluctuations, only the holes emit
and absorb magnons of the Cu magnetic background. For the Ising
$H_{J_{dd}}$, magnons are static flipped Cu spins. The absence of
dispersion as compared to a Heisenberg $H_{J_{dd}}$ may seem
problematic, but again it is the small $J_{dd}$ that controls the
magnon speed. Because this speed is small compared to that of other
relevant processes, it can be safely set to zero: a magnon emitted by
a hole is simply too slow to move away before it is absorbed either by
the same hole or by a different one. Another way to think about this
is that what matters here are real-space configurations, {\em i.e.}
how far is a magnon from a hole. A local (in space) magnon is a linear
superposition of all $\vec{q}$-momentum magnons. For symmetry reasons,
the coupling of small-$\vec{q}$ magnons to holes vanishes, so the holes
interact mostly with the large-$\vec{q}$ magnons, whose
dispersion is rather flat and which, therefore, can be safely treated as being immobile.

If only the holes create and absorb magnons, we can meaningfully
classify variational spaces in terms of their magnon numbers: the more
magnons, the higher their Ising energy cost, and the less likely to
find such configurations contributing significantly to low-energy
eigenstates. In this work, we limit the variational space to have up
to two magnons, and moreover require that any magnon is within a
distance $m_C$ from a hole - the reason being that we are interested
in the low-energy states where the magnons belong to {\em qp} clouds
and therefore are never too far from holes. This variational space
suffice to allow us to characterize the magnon-exchange between holes, which is
our goal. It is also sufficient to quantitatively capture the
dispersion of a single quasiparticle. \cite{HadiNP, HadiJCP} For two
holes, this space is too limited and overestimates the quasiparticles'
bandwidth, but this aspect can be mitigated (see discussion below).
The alternative of increasing the variational space (and thus run
times and memory resources) by allowing more magnons is less palatable considering that we
are already dealing with up to $10^6$ configurations. This is because
for two-hole configurations we need a second cutoff $M_C$ for the
maximum allowed distance between any two objects (holes and/or
magnons); to study unbound states properly, this cutoff can run to
many tens of lattice constants. More technical details regarding the
variational space, as well as the single-{\em qp} dispersion, are presented in Appendix \ref{app2}.

\begin{figure}[t]
  \centering \includegraphics[width=\columnwidth]{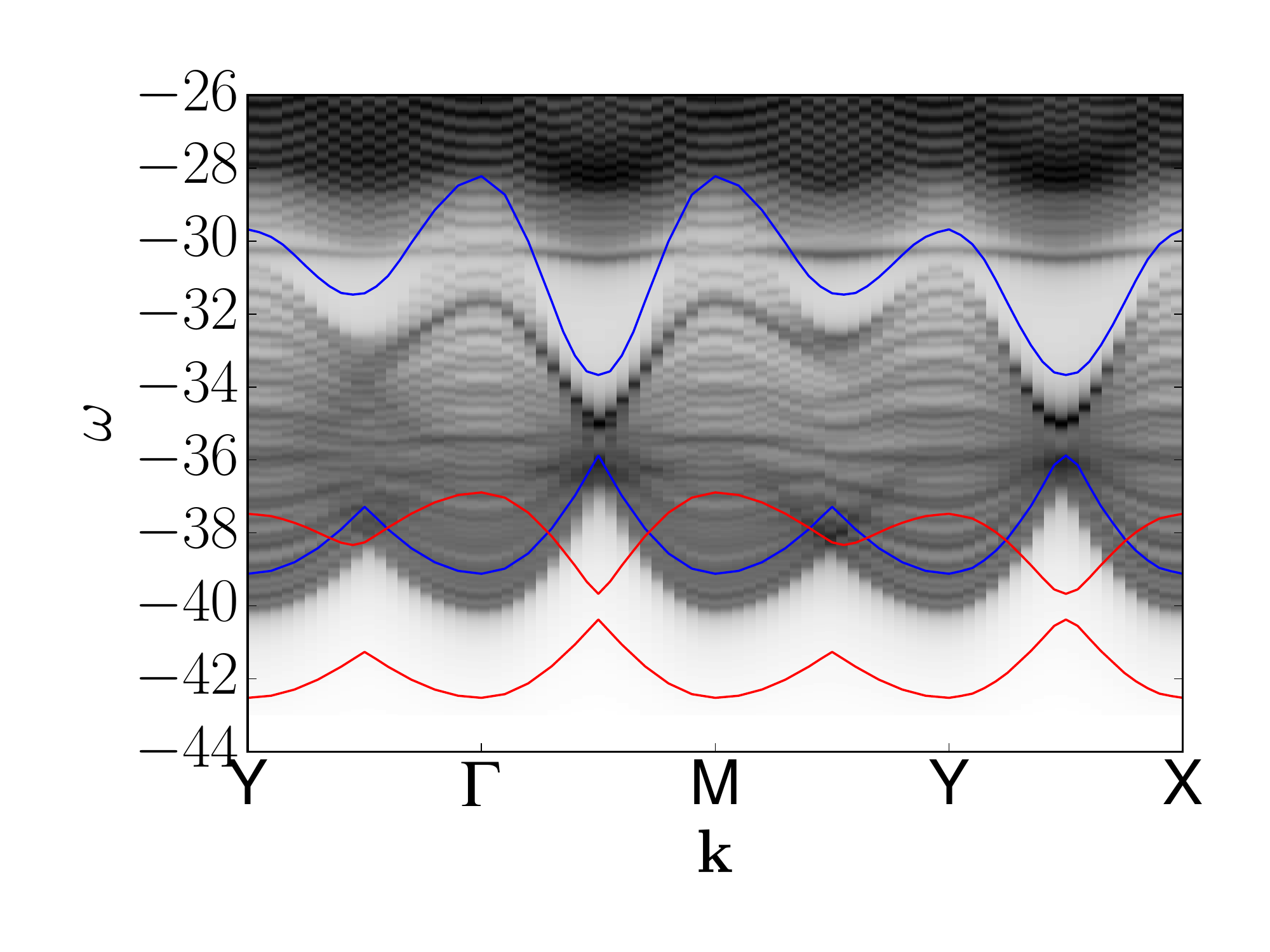}
  \caption{Contour plot of the two-hole spectral function
    $A(\mb{k},\omega)$ along the high symmetry
    lines of the BZ. This spectral weight is for states where one hole is on the $p_x$, and the other is on the $p_y$ orbital adjacent to the same Cu. Clearly, the  lowest-energy feature in the
    two-hole spectrum is a continuum. The blue and red lines indicate
    the expected continuum boundaries obtained from the convolution of
    single-{\em qp} dispersions when $n_m = 1$ and $n_m = 2$,
    respectively, {\em i.e.} when each {\em qp} is allowed to have up
    to 1 or up to 2 magnons in its cloud, respectively. The other
    cutoffs are $M_c = 40$ and $m_c = 3$.}
  \label{fig:S3}
\end{figure}

Diagonalizing Hamiltonian (1) in this variational space reveals that
the lowest feature of the two-hole spectrum is the continuum
describing two unbound {\em qps}, shown in Fig. \ref{fig:S3} as the
gray-scale contour plot. To verify this, we use our knowledge
of the single {\em qp} dispersion $\epsilon_{\mr{sp}}(\mb{k})$ (shown
in Appendix \ref{app2}) to find the expected location of the two-hole
continuum. This corresponds to the convolution of two single-{\em qp}
spectra, and for a total momentum $\mb{k}$ it spans $\{
\epsilon_{\mr{sp}}(\mb{k}-\mb{q}) + \epsilon_{\mr{sp}}(\mb{q})
\}_{\mb{q} \in \mr{BZ}}$. The blue (top) set of lines show the
expected location of the continuum if each {\em qp} cloud is
constrained to have up to $n_m=1$ magnons, while the red (bottom) set
of lines is the answer if each {\em qp} cloud is constrained to have
up to $n_m=2$ magnons.

As expected, our answer lies in between the two limits,
because in the two-hole variational space 
that we use, with some probability each {\em qp} can have more than one
magnon, however the space is not large enough so that both holes can have
2 magnons each at the same time. We verified that if we impose the
additional restriction for the two-hole variational space that when 2
magnons are present, each hole has a magnon within $m_c$ of it, we
recover perfect agreement between the two-hole continuum and the
$n_m=1$ single-hole prediction (not shown). We can also
artificially increase $J_{dd}$ leading to a higher energy cost for
magnons and thus less weight on the two-magnon states. For
very large $J_{dd}$, the red and blue lines in Fig.
\ref{fig:S3} fall on top of each other and coincide with the continuum
edge of the two-hole calculation (not shown).

We are thus confident that this lowest-energy continuum is indeed the
two-{\em qps} continuum, which is always a part of the two-hole
spectrum. Unfortunately, this result gives no clue about the
effective interaction between the {\em qps}, as the continuum would be
present whether the two {\em qps} are non-interacting or whether they
experience attraction or repulsion. All we can say is that {\em if}
there is magnon-mediated attraction between the two {\em qps}, it does not appear to be
strong enough to bind them into a ``pre-formed'' pair, which would
be a discrete state lying below this two-{\em qp} continuum (we
revisit this point below). However, from this result we cannot even
infer whether there is a magnon-mediated interaction.

To do that, we need to find a way to turn off magnon-exchange
processes, in order to gauge their effect on the two-{\em qps}
eigenstates. We describe how we achieve this goal in the next section.

\section{Quantifying the effective magnon-mediated interaction}

\begin{figure}[t]
\includegraphics[width=\columnwidth]{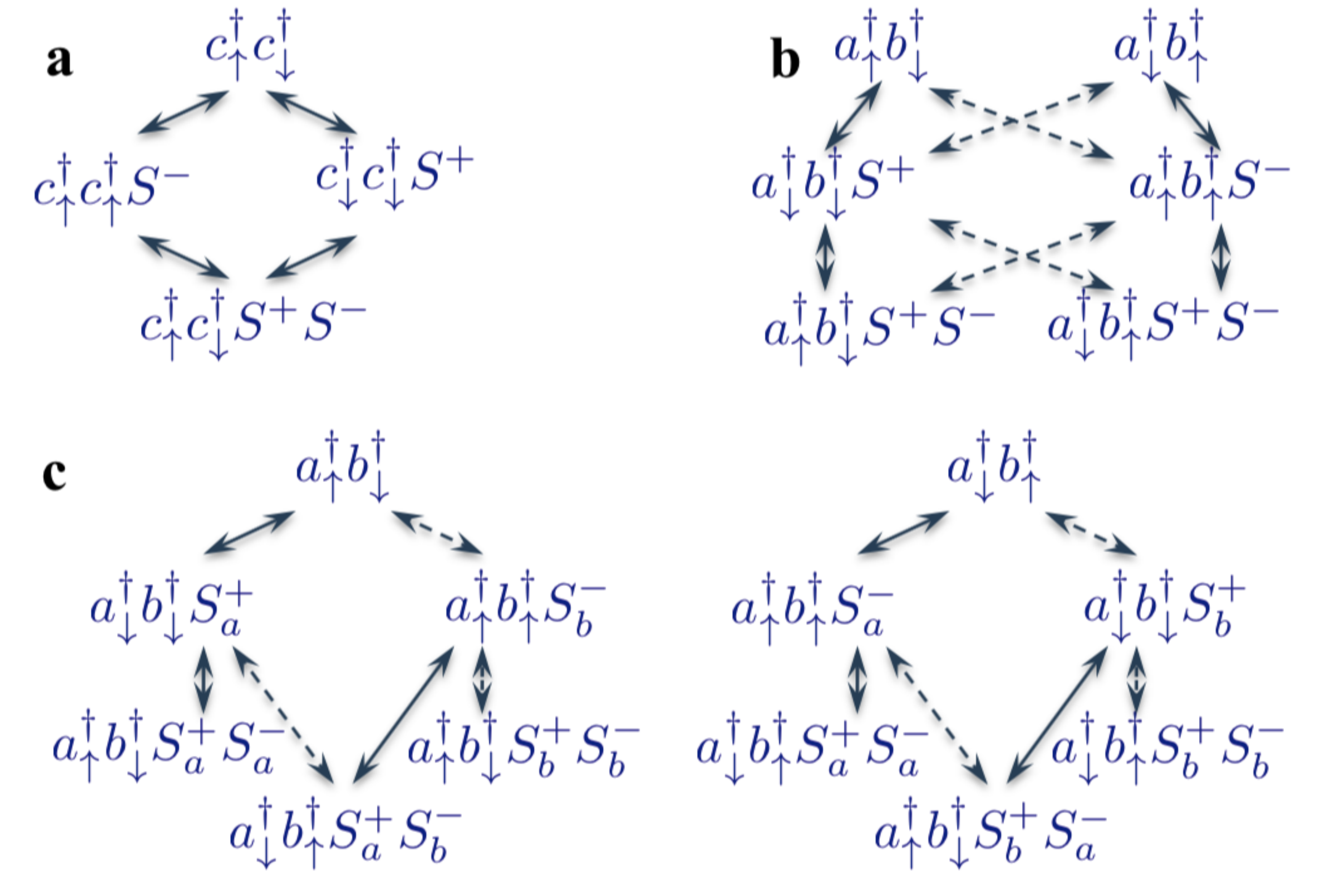}
\caption{
  (a) Original variational space, consisting of no-magnon (top line),
  one-magnon (middle line) and two-magnon (bottom line) configuration.
  The holes and/or magnons can be at any sites consistent with the
  cutoffs, but for simplicity we do not label their positions; (b)
  Variational space when the holes are given  flavors $a$ or $b$. This
  can be mapped exactly onto the original variational space of (a)
  using antisymmetric combinations of the two flavors; (c) Extended
  variational space where magnons also have a flavor. This allows us
  to turn off the exchange of magnons between holes.}
\label{fig1b}
\end{figure}

As already mentioned in the Introduction, the key difficulty with
turning off magnon-exchange processes is in figuring out when a magnon
has actually been exchanged. This can be seen by considering the
configurations of the variational space, sketched in Fig.
\ref{fig1b}(a). The top line indicates configurations with a spin-up
and a spin-down hole plus the AFM background (the holes are at various
locations but for simplicity we do not label these). Either hole can
create a magnon in the appropriate magnetic sublattice; this
leads to the one-magnon configurations from the second line. Either
hole can emit a second magnon resulting into two-magnon configurations
like in the third line. In principle, any number of magnons can be
emitted so this hierarchy of configurations is infinite, but as
mentioned we keep only up to two-magnon configurations in our
variational space.

In the zero- and two-magnon configurations, the holes are
distinguishable through their spins (no term in Hamiltonian (1) allows
direct hole-hole spin exchange). However, in the one-magnon
configurations both holes have identical spin and thus are
indistinguishable. This is why when considering the magnon absorption
from such a configuration, it is impossible to know which of the two
holes flipped its spin to emit the magnon in the first place. As a
result, we cannot forbid magnon-exchange processes at this level, as
this requires us being able to distinguish between the indistinguishable
holes.

We therefore must assign different flavors $a$ and $b$ to the holes, so that they
are distinguishable even when they have the same spin. This results in
the configurations of Fig. \ref{fig1b}b. Interestingly, these two
variational spaces map exactly onto each one if we use the correspondence
$c^\dagger_{\sigma}c^\dagger_{\sigma'} \leftrightarrow
(a^\dagger_{\sigma}b^\dagger_{\sigma'}-
a^\dagger_{\sigma'}b^\dagger_{\sigma})/\sqrt{2}$, which is necessary
to enforce Pauli's principle. However, for these antisymmetrized,
physical states, it is still impossible to know which particle emitted
the magnon, just like for the original states onto which they map. For
instance, if the system is in a $a^\dagger_{\downarrow}
b^\dagger_{\downarrow} S^+$ type of configuration, it may have arrived
there either by starting in the $a^\dagger_{\uparrow}
b^\dagger_{\downarrow}$ sector of the physical state with the $a$-type
hole emitting the magnon, or by starting in the $a^\dagger_{\downarrow}
b^\dagger_{\uparrow}$ sector with the $b$-type hole emitting the
magnon. The two scenarios cannot be distinguished, therefore we still
cannot know which hole emitted the magnon so we cannot decide if a
magnon-absorption process is of magnon-exchange type or of
quasiparticle renormalization type.

This is why we need to also label the magnons as $a$ or $b$, according
to which hole emitted them. This leads to the variational space
sketched in Fig. \ref{fig1b}c, which we call the ``enlarged
variational space''. In this enlarged space we can turn off the
magnon-exchange by requiring that an $a/b$ magnon can only be absorbed
by an $a/b$ hole. Its states are divided into two families that do not
mix if the magnon-exchange is turned off, which is the situation
sketched in Fig. \ref{fig1b}c. Again, the physical states are the
antisymmetrized combinations originating from
$(a^\dagger_{\sigma}b^\dagger_{\sigma'}-
a^\dagger_{\sigma'}b^\dagger_{\sigma})/\sqrt{2}$ zero-magnon
configurations, but now for the situation sketched in Fig.
\ref{fig1b}c, we know that to arrive at a $a^\dagger_{\downarrow}
b^\dagger_{\downarrow} S^+_a$ configuration, the $a$-type of hole
emitted the magnon. If only the $a$-type hole can absorb it, then
magnon-exchange is turned off, and we can contrast the results in its
absence to those obtained when magnon-exchange is allowed. From such
comparisons we ought to be able to infer the effects of the
magnon-exchange, in particular whether they result in an effective
attraction between the {\em qps}.

Before continuing, we must note that the mapping of the antisymmetrized,
physical states from the extended variational space onto their
counterparts in the original basis is no longer one-to-one, because of
the increased number of one- and two-magnon configurations. Instead,
the enlarged variational space can be thought of as corresponding to
the tensor product of the variational spaces for single spin-up and
spin-down holes, respectively, but with the physical constraints
imposed, {\em e.g.} two magnons cannot be at the same Cu site, etc. This
is meaningful because in the low-energy states, magnons are bound to
their hole's cloud and therefore need not be treated as free
particles.

\begin{figure}[t]
  \includegraphics[width=\columnwidth]{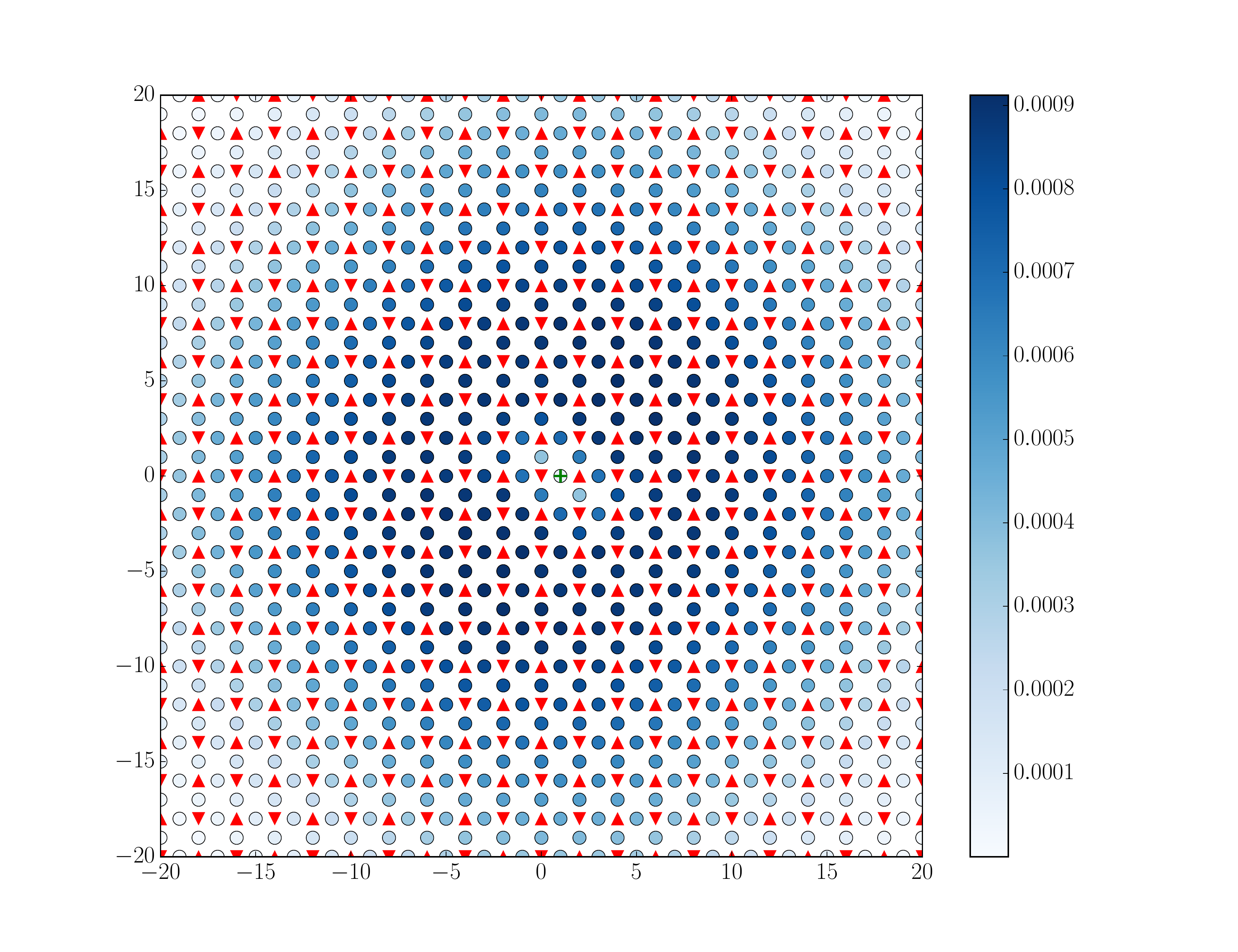}
  \includegraphics[width=\columnwidth]{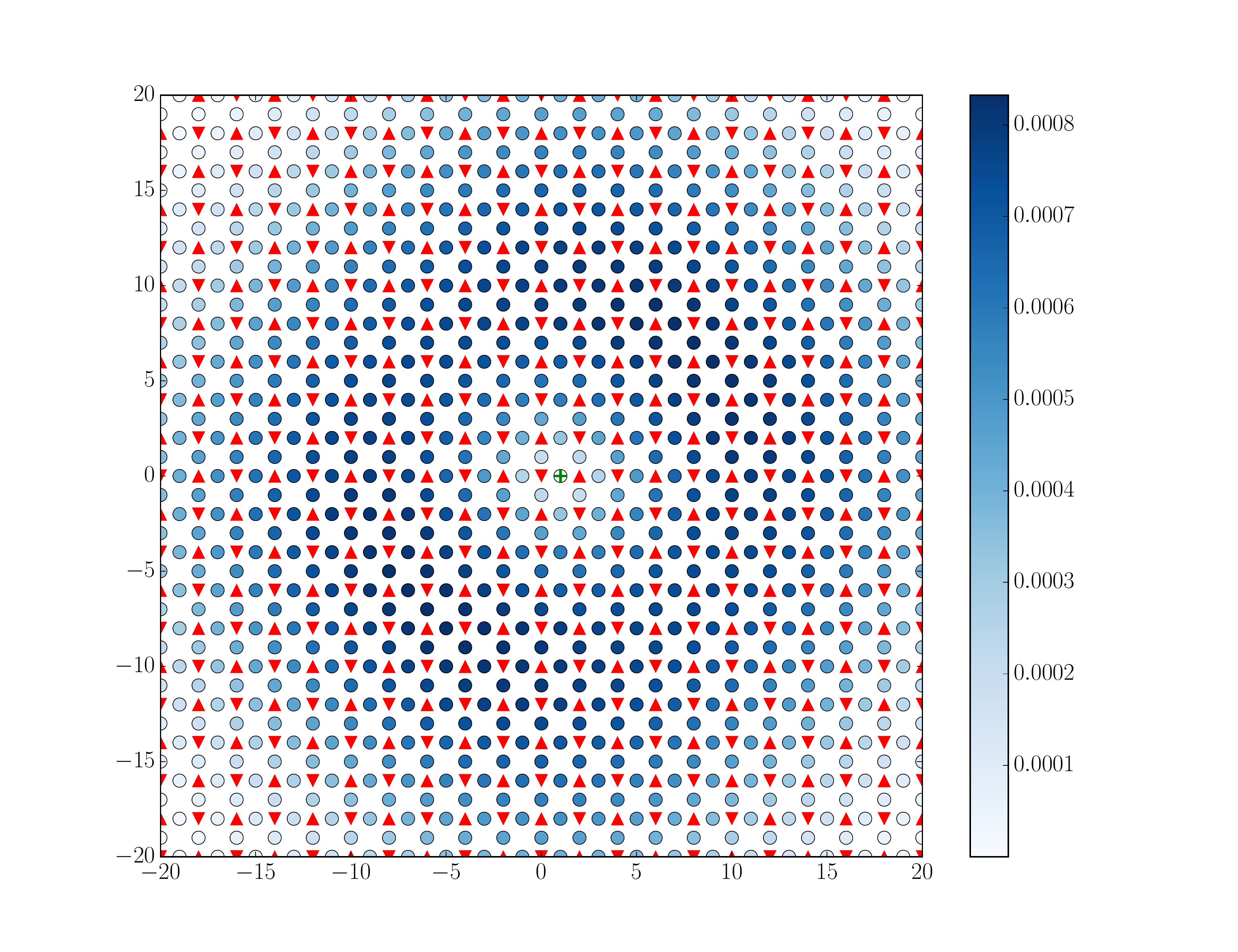}
\caption{Contour plots (shades of blue) for the GS probability to find
  a hole on various O sites (shown by circles), if the other other
  hole is at the central $p_x$ O site marked with the green cross. The
  red arrows are at the positions of Cu and indicate the direction of
  the spins {\em in the undoped ground state}, {\em i.e.} before the
  holes were added. Top panel shows the result when magnon-exchange is
  allowed, while the bottom panel shows the result when
  magnon-exchange is turned off. }
\label{figcont}
\end{figure}

Figure \ref{figcont} shows GS results for the case when the
magnon-exchange is allowed (top panel) {em vs.} forbidden (bottom
panel). The red arrows are at the positions of Cu sites and indicate the
direction of their spins {\em in the undoped ground state}. Of course,
the spin order is modified by the presence of the holes but showing
that in a meaningful way on this scale is impossible, which is why we
show the magnetic order before the holes were introduced. The O
locations are indicated by circles. Their blue shading 
indicates the probability to find a hole on that O, if the other hole
is located on the central $p_x$ orbital marked by the green cross.

Clearly, when the magnon-exchange is allowed, the the holes are closer
than when the magnon-exchange is turned off. This clearly proves that
{ \em magnon-exchange mediates an effective attraction between the two
  holes}. This is one of the main results of this study.

\begin{figure}[t]
  \includegraphics[width=0.6\columnwidth]{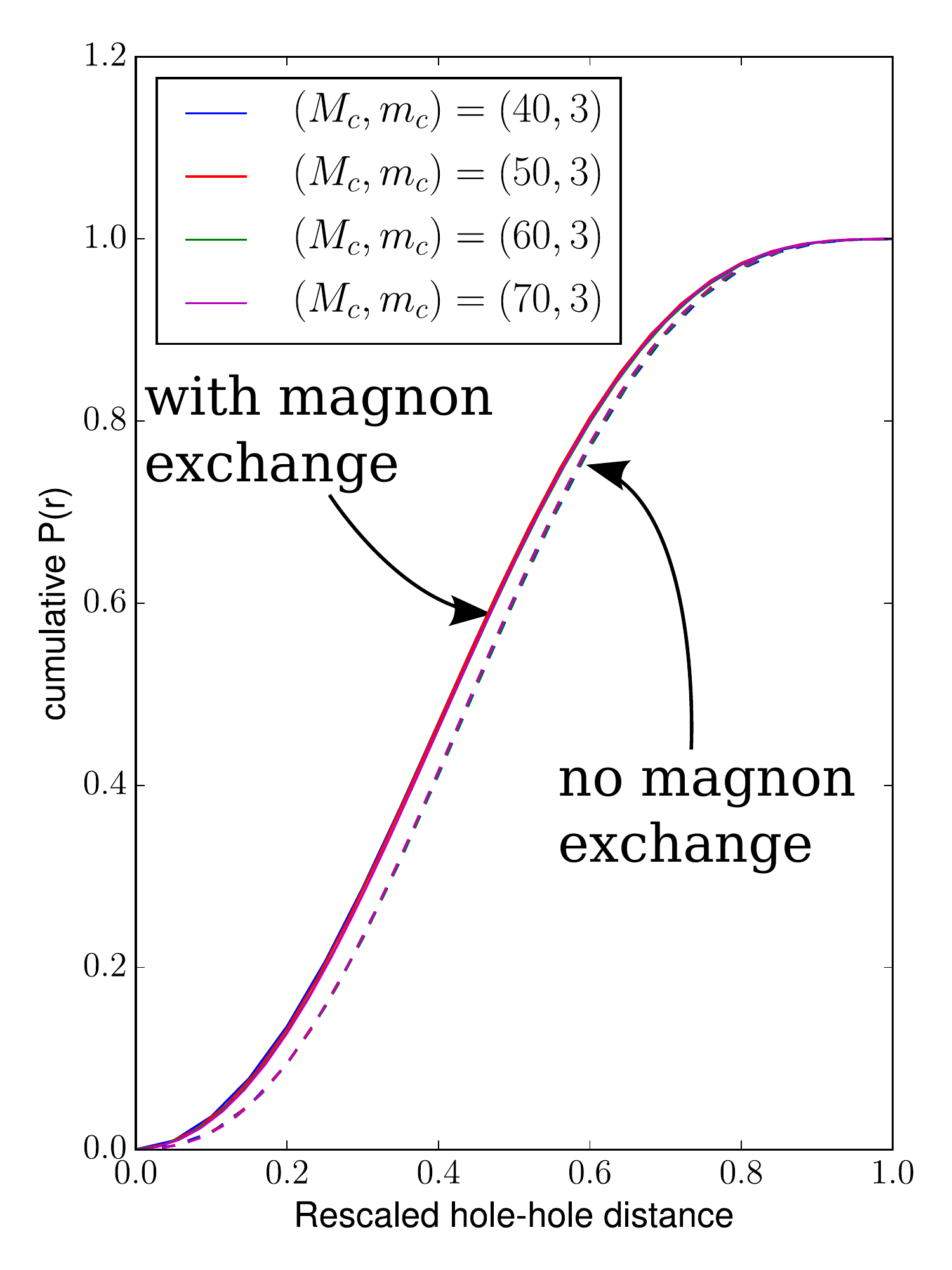}
\caption{(color online) GS cumulative probability $P(r)$ to find the two holes
  within a distance $r/M_c$ of each other, when magnon-exchange is
  allowed/forbidden (full/dashed lines). The results are essentially independent
  of the values of the cutoffs $M_c, m_c$. }
\label{figPvsM}
\end{figure}

Before trying to quantify how strong is this attraction, we point out
two important facts. First, this GS is at the bottom of the two-{\em
  qps} continuum, so these two holes are not bound. The reason why there
is finite probability for them to be close to one another is the
existence of the constraint $M_c$ on the largest relative distance
allowed between them. This imposes a ``finite-box'' type of
restriction on the relative motion of the two holes, so they cannot
move infinitely far apart. We have checked that the tendency of holes
to be closer when magnon-exchange is allowed is independent of the
size of the $M_c$ cutoff, as indeed expected for an interaction with a
finite range. This is shown in Figure \ref{figPvsM}(a), where we plot
the cumulative probability $P(r)={1\over N} \sum_{i}^{}
\sum_{|j|<r}^{} \langle GS| \hat{n}_i\hat{n}_{i+j}|GS\rangle$ to find
the holes within a distance $|j|\le r$ (measured using the L1-norm)
versus the scaled separation $r/M_C$, for several values of the
maximum allowed relative distance $M_C$. Curves with different $M_C$
overlap, showing that these are indeed unbound states: the holes move
further apart if $M_C$ is larger. The full/dashed lines show the
results with/without magnon-exchange. In its presence the holes are
closer to each other, therefore magnon-exchange mediates an effective
attraction between holes.

The second note is that the GS is doubly degenerate. The
contour plots of Fig. \ref{figcont} will look somewhat different
depending on which linear combination of the two eigenstates is chosen
for calculating the probability. The choice we made in Fig.
\ref{figcont} is to use the eigenstate that is even to reflections
about the $x-y$ diagonal. Irrespective of which choice is made, the
holes are always closer to one another when magnon-exchange is allowed.

 Next, we identify $H_{\rm{eff}}$ that describes this magnon-mediated
 attraction. This is achieved by adding various possible candidates
 for $H_{\rm{eff}}$ to the calculation in the enlarged space without
 magnon-exchange, and adjusting until the results match those with
 magnon-exchange allowed. We use perturbation theory (PT) to suggest
 possible forms: $H_{\rm{eff}}\sim \hat{P}_0 \hat{V} \frac{1- \hat{P}_0}
 { E_0-{\cal H}_0} \hat{V} \hat{P}_0 $, where $\hat{P}_0$ projects
 onto the zero-magnon subspace, ${\cal H}_0$ contains terms that
 conserve the number of magnons and $\hat{V}= {\cal H}-{\cal H}_0$
 contains terms that create or annihilate magnons.

 Clearly, ${\hat V}$ has contributions from $T_{sw}$ and $H_{J_{pd}}$.
 These are further divided into direct processes where holes
 create/remove magnons of their own flavor, {\em vs.} exchange ones,
 where they interact with magnons of the other flavor (only
 direct processes are allowed when the magnon-exchange is turned off).
 To mimic the effect of the magnon-exchange, $H_{\rm{eff}}$ must
 contain the product between a direct and an exchange term --
 {\em eg.} a hole emits a magnon of its own flavor (direct process)
 which the other hole then absorbs (exchange process).

 Generating all such terms suggested by PT and
 enforcing hermiticity, we find four possible contributions to
 $H_{\rm{eff}}$ (the details 
 are provided in Appendix \ref{app3}). We then investigate each
 term separately, treating its magnitude as a free parameter, fitted to get
 the same GS energy as for the full calculation with magnon-exchange
 allowed. This approach allows us to account for the renormalization
 of this energy scale due to higher order terms in the perturbative
 expansion.
 
We find that the dominant term has both the magnon emission and 
absorption due to $T_{\rm{sw}}$. This finding is consistent with the
independent check that turning off the magnon-exchange due to this term accounts for nearly all
the difference between the cumulative probabilities of Fig.
\ref{figPvsM} (not shown).

\begin{figure}[t]
\includegraphics[width=0.9\columnwidth]{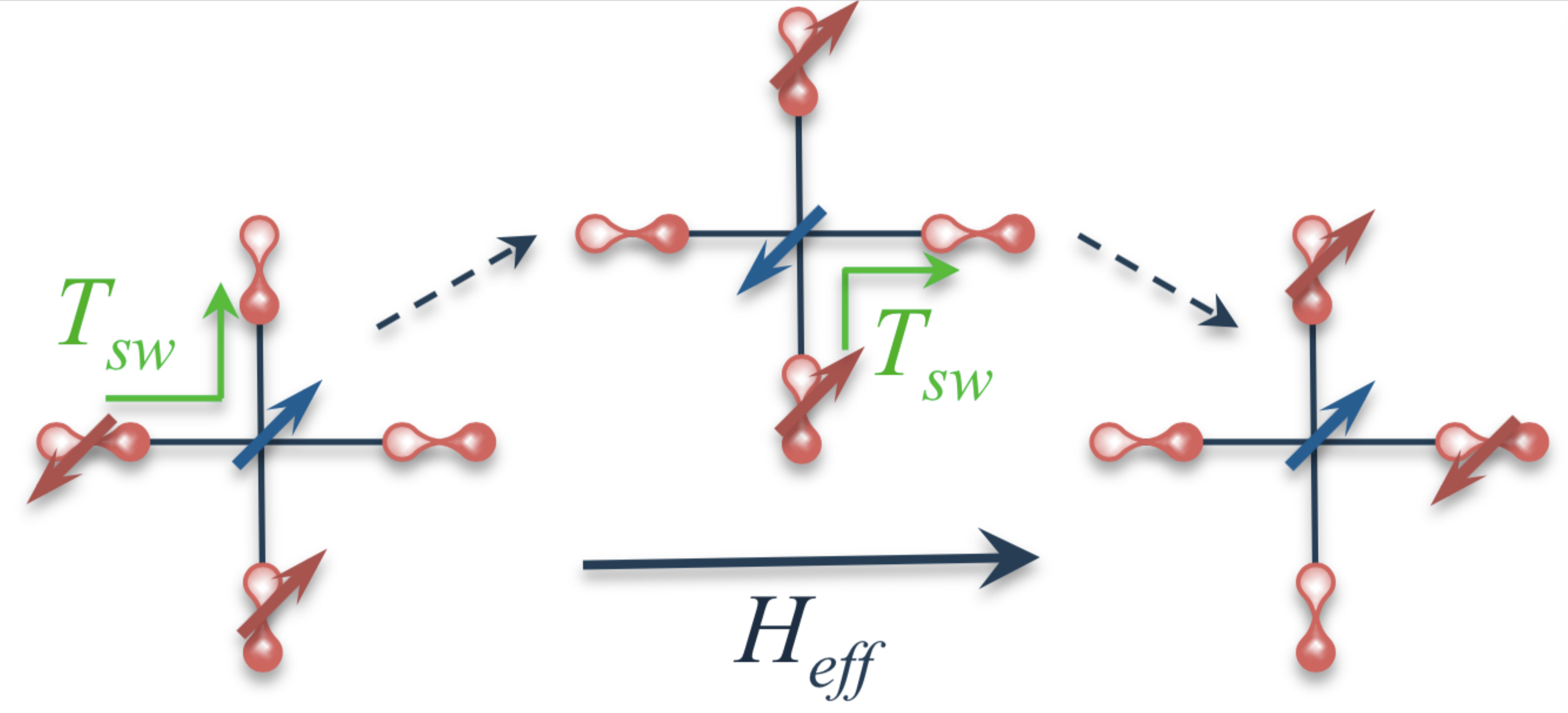}
\caption{(color online) Sketch of one of the terms in $H_{\rm eff}$. Both holes hop through $T_{\rm{sw}}$ processes, exchanging a magnon through the common Cu spin. All possible such processes are included in  $H_{\rm eff}$, with the same magnitude but signs depending on $pd$ overlaps.}
\label{fig6}
\end{figure}

Keeping only this dominant term, we find
$H_{\rm{eff}}= \sum_{j } (H_{\rm{eff}}^{j\uparrow} + H_{\rm{eff}}^{j\downarrow} )$ where the
two terms correspond to having both holes adjacent to the up/down Cu
spin in the unit cell $j$, and:
\begin{align}
  \label{eq:Heff-analytical}
  & H_{\rm{eff}}^{j\uparrow} = t_{\rm{pair}}\!\!\! \!\!\!
  \sum_{\alpha,\beta, \eta_\alpha\atop\eta_\beta \neq u_\alpha -
    u_\beta } \!\!\! \!\!\! \zeta_{\alpha\beta}
  c_{j-u_\beta - \eta_\beta, \uparrow}^\dagger c_{j-u_\alpha
    -\eta_\alpha, \downarrow}^\dagger c_{j-u_\beta ,
    \downarrow}c_{j-u_\alpha, \uparrow} \nonumber \\ &
  H_{\rm{eff}}^{j\downarrow} = t_{\rm{pair}}\!\!\!
  \!\!\!\sum_{\alpha,\beta,\eta_\beta\atop \eta_\alpha \neq u_\beta -
    u_\alpha}\!\!\! \!\!\! \zeta_{\alpha\beta}
  c_{j+u_\beta + \eta_\beta, \uparrow}^\dagger c_{j+u_\alpha +
    \eta_\alpha, \downarrow}^\dagger c_{j+u_\beta ,
    \downarrow}c_{j+u_\alpha, \uparrow} \nonumber
\end{align}
In units of ${a\over 2}$, the vectors ${\bf u}=\pm (1 ,0), \pm (0, 1)$ show
the locations of O neighboring the Cu spin, while the vectors ${\bf \eta}=\pm(1,
1), \pm(1, -1), \pm( 2,0), \pm(0, 2)$ link O sites adjacent to the
same Cu. Finally, $\zeta_{\alpha\beta}=-2
\rm{s}(\eta_\alpha) \rm{s}(\eta_\beta)$, where
$\rm{s}(\eta)=+1$ if $\eta_x+\eta_y=0$, otherwise
$\rm{s}(\eta)=-1$.

This Hamiltonian is the {\em main result} of this work. It contains
conceptually simple processes like that sketched in Fig. \ref{fig6}:
 First, the hole with spin antiparallel to the common Cu
spin undergoes a $T_{\rm{sw}}$ process and moves to another O while
swapping its spin with the Cu. This amounts to the emission of a
magnon at that Cu site, subsequently absorbed when the second hole
undergoes a $T_{\rm{sw}}$ process involving the same Cu. Thus,
$H_{\rm{eff}}$ describes both holes hopping as a pair but also
exchanging their spins. The relative signs $\zeta_{\alpha\beta}$ are due to the
product of appropriate $T_{\rm{sw}}$ signs, which in turn are
controlled by the overlaps between the O$2p$ and Cu$3d$ orbitals
involved. \cite{Bayo}

\begin{figure}[t]
\includegraphics[width=\columnwidth]{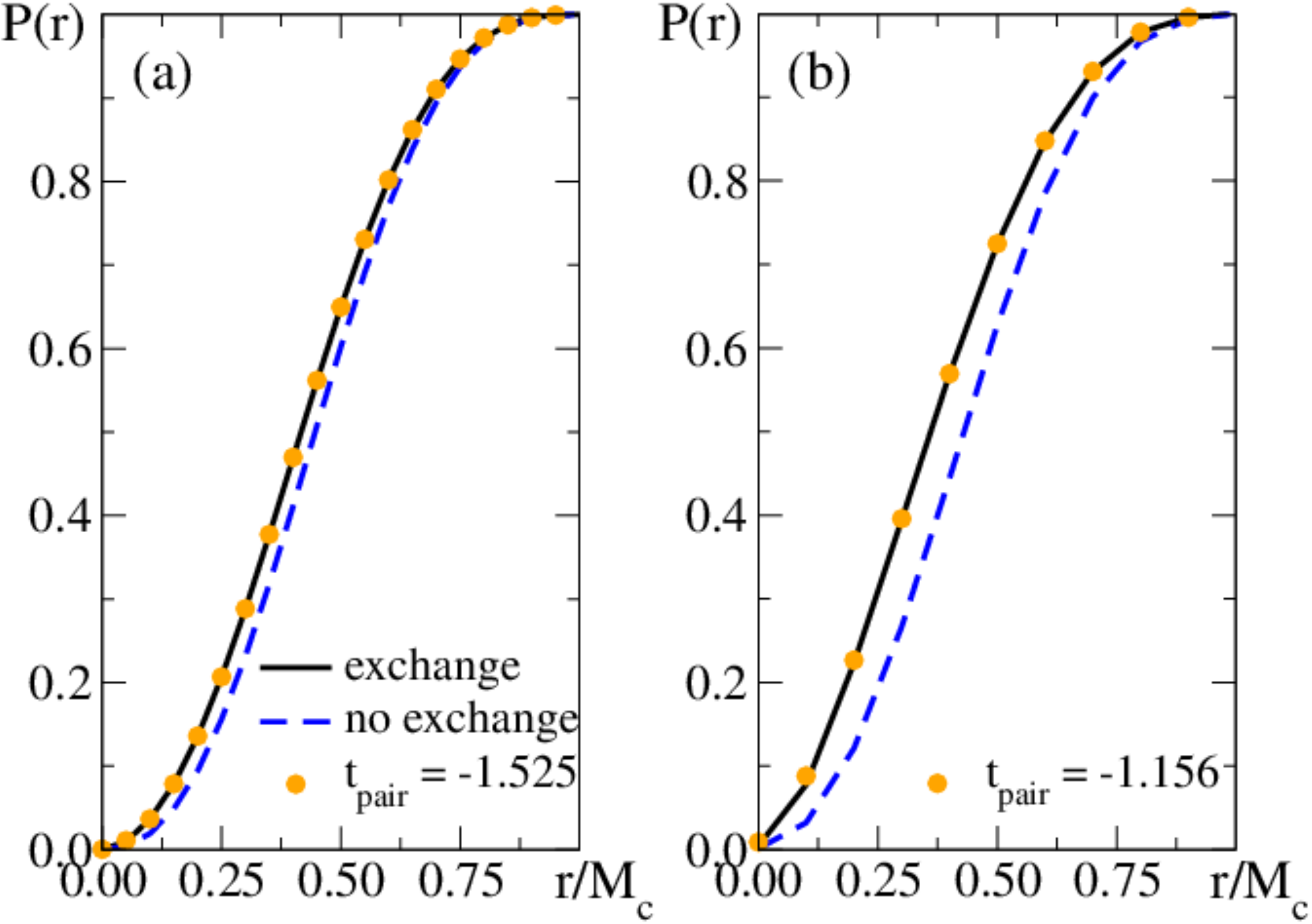}
\caption{(color online) (a) GS cumulative probability $P(r)$ to find the two holes within a distance $r/M_c$ of   each other, when magnon-exchange is allowed/forbidden (full/dashed lines). These are the same results as in Fig. \ref{figPvsM}. In addition, the symbols show $P(r)$ when the magnon-exchange is turned off but $H_{\rm eff}$ is added instead, with a $t_{\rm pair}=-1.525$. (b) Same as in (a) but for $U_{pp}=0$. In this case, $t_{\rm pair}=-1.156$. }
\label{fig7}
\end{figure}

\begin{figure}[t]
\includegraphics[width=\columnwidth]{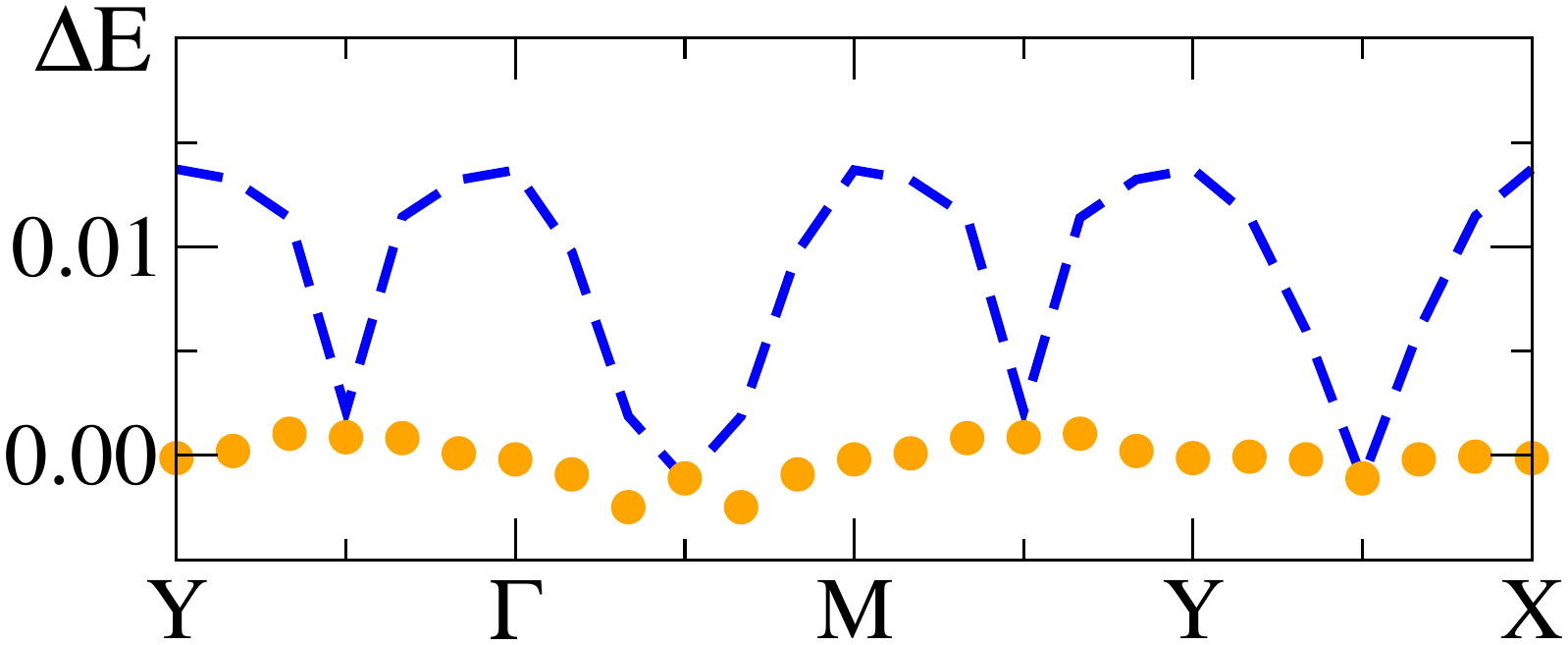}
\caption{(color online) Differences 
between the lowest eigenenergies without and with magnon-exchange
(dashed line) versus the same difference but with $H_{\rm{eff}}$
included when the magnon-exchange is turned-off (symbols). }
\label{fig8}
\end{figure}

We find that adding this $H_{\rm{eff}}$ to the calculation with
magnon-exchange forbidden produces the same GS energy as the full
calculation with the magnon-exchange allowed if we set
$t_{\rm{pair}}=-1.525$. To validate it, in Fig. \ref{fig7}a we show
that the GS cumulative probability in the enlarged space without
magnon-exchange but with this $H_{\rm{eff}}$ included (symbols)
matches perfectly that obtained when magnon-exchange is allowed (full
line). This shows that this $H_{\rm{eff}}$ reproduces the GS
wavefunction accurately. Furthermore, we find that it gives a faithful
description of the effects of magnon-exchange in the {\em entire}
Brillouin zone: Fig. \ref{fig8} compares the differences $\Delta
E_{\rm no-ex} = E_{\rm no-ex}-E_{\rm ex}$ and $\Delta E_{H_{\rm eff}}
= E_{H_{\rm eff}}-E_{\rm ex}$ between the lowest eigenenergies without
and with magnon-exchange (dashed line) versus the same difference but
with $H_{\rm{eff}}$ included if magnon-exchange is forbidden
(symbols). Note that $E_{ex}$ is shown in Fig. \ref{fig:S3}, as the
lowest energy for each momentum in the Brillouin zone.

Clearly,
$H_{\rm{eff}}$ reproduces very well the effect of the magnon-exchange
in the full Brillouin zone, even though $t_{\rm
  pair}$ is fitted only for agreement at the $\Gamma$ point. Note that these
energy variations are again due to the finite $M_c$ constraint. While
the size of the  energy differences depends on the value of $M_c$, we verified
that adding this  $H_{\rm{eff}}$ works well for any value of $M_c$.

We therefore conclude that this $H_{\rm{eff}}$ indeed reproduces very
well the effect of the magnon-mediated attraction between the two
holes. This is a non-trivial result, as there is no {\em a priori}
reason to expect that $H_{\rm{eff}}$ contains a single class of
processes. As mentioned, 2$^{nd}$ order PT suggests three other
possible candidates, involving $H_{J_{pd}}$ in the magnon emission
and/or absorption. Although $H_{J_{pd}}$ and $T_{\rm{sw}}$ have
comparable energy scales, these other processes turn out to have
little effect, in other words higher order PT terms seem to renormalize them
to become vanishingly small.
We do not currently have a good understanding as to why this happens.

Of course, higher order PT terms also generate other possible exchange
scenarios, involving more magnons. That these do not contribute much
is less surprising because all their magnons have to be exchanged,
{\em i.e.} created by a hole and absorbed by the other in a way that
is not just a sequence of independent $H_{\rm{eff}}$ processes. This
is rather difficult given the structure of the CuO$_2$ planes, which
makes it easy for two holes to be neighbors of the same Cu spin, but
impossible to be simulataneously neigbors of two or more different Cu
spins.

The expression of this effective, magnon-mediated attraction
$H_{\rm{eff}}$ is the central result of this work. It is very
different from the more customary density-density or exchange type of
effective interactions previously used in the literature. As such, it is
likely to drive different behaviour at higher
concentrations; investigation of these differences is left for future
work.

\begin{figure}[t]
\includegraphics[width=0.79\columnwidth]{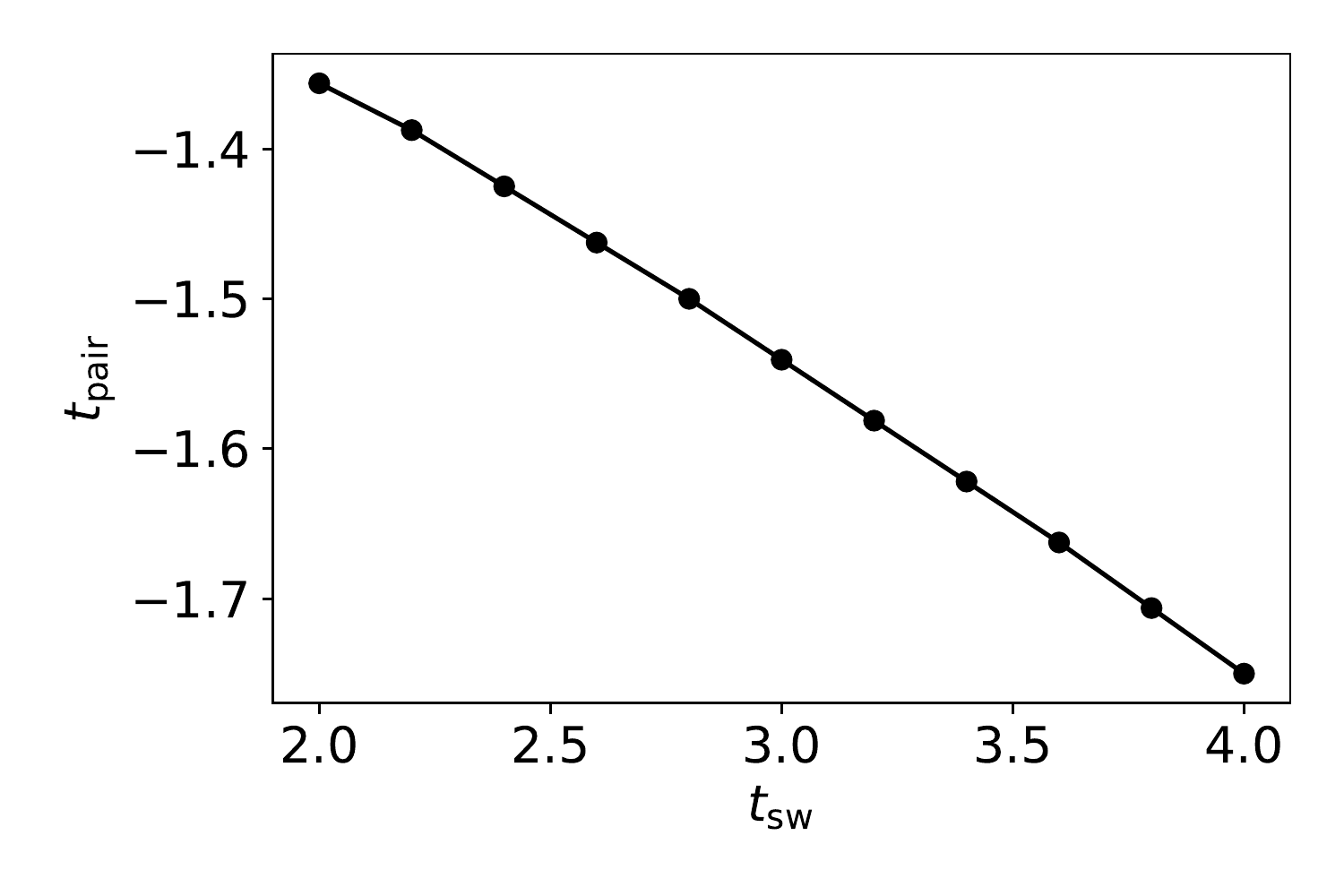}
\includegraphics[width=0.81\columnwidth]{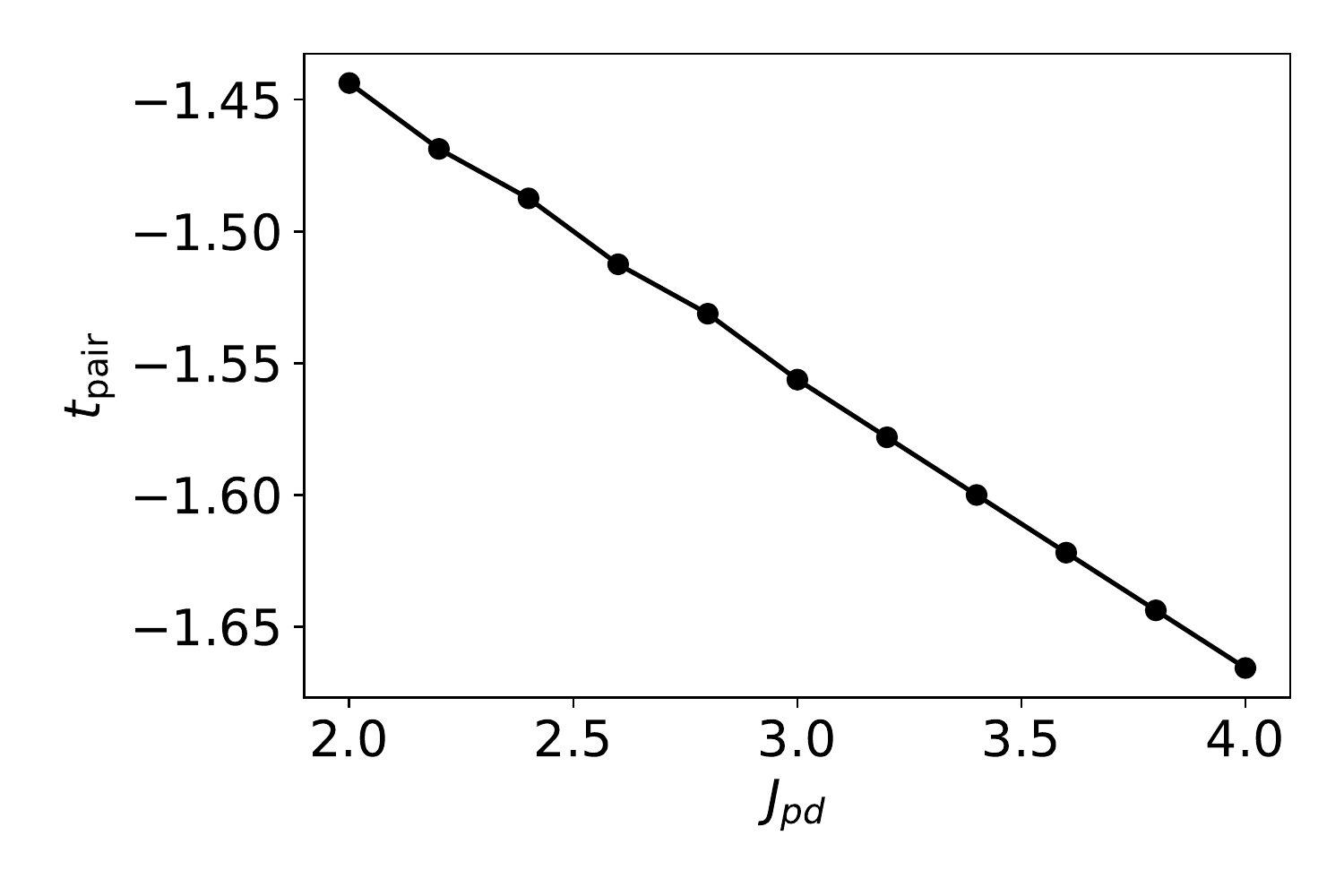}
\includegraphics[width=0.74\columnwidth]{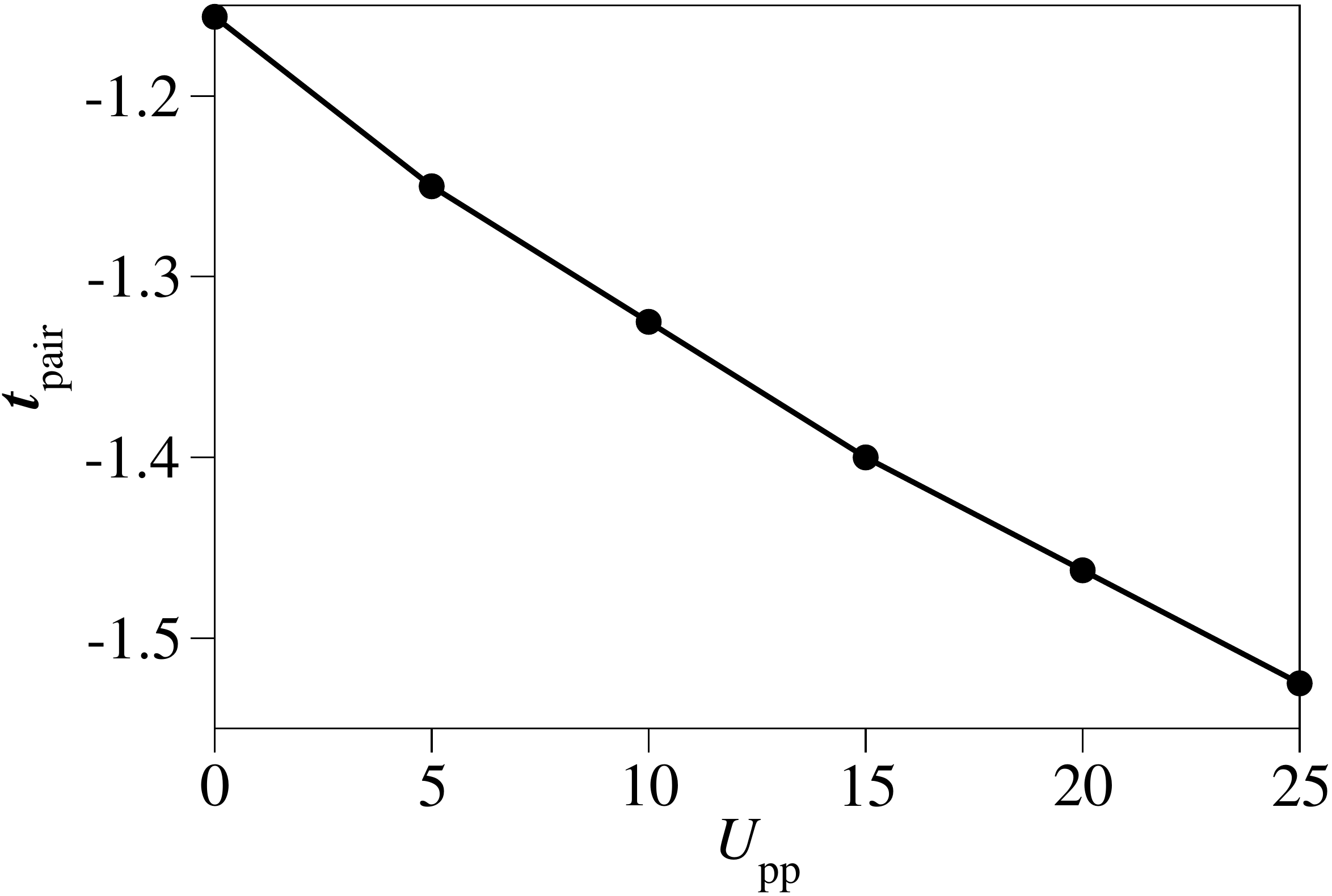}
\caption{Dependence of $t_{\rm pair}$ on $t_{\rm sw}$ (top), on $J_{\rm pd}$ (middle) and $U_{pp}$ (bottom). In each case, all other parameters are held fixed at their stated values. }
\label{fig9}
\end{figure}

Before concluding this section, we briefly address the dependence of
$t_{\rm{pair}}$ on the various parameters. Most is as expected, {\em
  eg.} monotonic increase with both  $t_{sw}$ and 
$J_{pd}$, shown in Figs.  \ref{fig9}a,b respectively. The surprise is that $|t_{\rm{pair}}|$ increases with
$U_{pp}$, see Fig. \ref{fig9}c. A larger $U_{pp}$ disfavors
configurations with both holes on the same O, thus fewer
pair-hopping+exchange processes are effectively allowed. Thus, it is
not obvious whether the larger $|t_{\rm{pair}}|$ value really means
stronger attraction because, at the same time, some terms in $H_{\rm
  eff}$ are effectively blocked. Indeed, the change in the cumulative
probability with and without magnon-exchange is much more significant
for $U_{pp}=0$ than that shown in Fig. \ref{fig7}b, suggesting that
the magnon-mediated attraction is stronger for smaller $U_{pp}$. This
serves to illustrate the fact that interactions like this $H_{\rm
  eff}$ have not been thoroughly studied and we lack intuition about
their effects.

\section{The role of background spin fluctuations}

We have repeated the analysis described above in the case when spin-fluctuations
(which allow any two adjacent, antiparallel Cu spins to simultaneouly
flip their spins) are allowed within $m_C$ of the holes. This
restriction is sensible because spin-fluctuations which occur far from
the holes can be thought of as ``vacuum fluctuations'' with which the
holes do not interact and which, therefore, will have no effect
either on the holes' dynamics or on the effective interaction between them.

In Ref. \onlinecite{HadiJCP} we proved that in the one-hole sector, this approach provides excellent
agreement with Exact Diagonalization (which fully includes the effects of spin fluctuations) both for our three-band
model, and for $t-t'-t''-J$ one-band models. We also showed
that spin fluctuations have no influence on the quasiparticle
(spin-polaron) dynamics in the three-band model, even though they play
an essential role in the one-band models. This is because as mentioned, in the
three-band model $J_{dd}$ (which defines the characteristic timescale
for spin-fluctuations) is significantly smaller than the other
energy scales, whereas its counterpart $J$ in the one-band models is
comparable to $t', t''$ (for more analysis, see Ref.
\onlinecite{HadiJCP}).

Redoing the two-hole calculation in the presence of local
spin-fluctuations reveals results very similar to those already
discussed (not shown). In particular, the value of the best fit for
$|t_{\rm pair}|$ varies by less that $5\%$. This clearly shows that
the spin-fluctuations play little role in the effective
attraction mediated by magnon-exchange, and validates our assertion that we can indeed ignore them.

This result is not surprising. As already mentioned several times,
$J_{dd}$ is the smallest energy scale, meaning that spin-fluctuations
happen on a very long (slow) time-scale. Roughly put, a magnon will be
exchanged between holes much faster than the timescale over which spin
fluctuations occur, which is why we can ignore them.

The same conclusion is reached if one tries to infer how
spin-fluctuations might affect the magnon-exchange process. Suppose,
for instance, that one of the holes flips its spin through either
$H_{J_{pd}}$ or $T_{\rm sw}$ and creates a magnon at the Cu site
denoted as '1'. If this is the first magnon, then spin '1' is now
parallel to its 4 Cu neighbors and spin-fluctuations cannot directly
act on it. Spin fluctuations could flip another pair of neighboring
spins (called '2' and '3') and then another spin fluctuation could
flip two of them (eg, '1' and '2') back to their original
orientations, leaving the magnon at site '3' (this sequence of events
basically mimics magnon dispersion). The magnon can now be absorbed by
the second hole. Clearly, this is a lot more complicated and less
likely (for a small $J_{dd}$) than the simple process where one hole
creates the magnon and the second absorbs it without further
complications.

A simpler scenario is when there already exists a magnon on a Cu site
neighbor to site '1', enabling a spin fluctuation involving spin '1'
to occur without further complications. However, this will remove both magnons, so ``magnon
exchange'' as normally envisioned would not occur. Nevertheless, if
the other magnon was emitted by the other hole, then this process
(and its counterpart, wherein a spin fluctuation creates two magnons,
each of which is absorbed by a different hole) will contribute to the
effective hole-hole interaction. Such processes are included in our
calculation when local spin fluctuations are allowed and, as
mentioned, were found to have only a tiny effect on the value of
$t_{pair}$.

\section{On the existence of preformed pairs}

So far, we found that for reasonable values of the parameters, the
magnon-mediated interaction does not appear to be strong enough to
bind the two holes into a preformed pair. To see how far away that
regime is, we can increase $t_{\rm pair}$ by hand (thus mimicking a
stronger magnon-mediated attraction) to find when binding occurs. An
exact answer is difficult to obtain because of the finite maximum
distance $M_c$ imposed between holes, which reduces the two-hole
continuum to a fairly dense sequence of discrete levels. The value of
$t_{\rm pair}$ where one state is pushed below this ``continuum''
changes somewhat with $M_C$, but we estimate that $|t_{\rm pair}|\sim
2.7$ is a safe upper limit -- for this and larger values of $|t_{\rm pair}|$ the energy
of the bound state is independent of $M_c$, and clearly well below the
continuum.

Thus, the value $|t_{\rm pair}|=1.525$ we found  is less than a factor of two from this
critical value. This is very interesting because the rather small
number of magnons kept in the variational space means that we
overestimate the quasiparticle bandwidth by about the same factor, as
shown in Figure \ref{fig:S3} (the red lines are essentially converged,
and show a significantly narrower bandwidth than the numerical
results). Given that binding occurs when the lost kinetic energy is
compensated by the increased attraction, this suggests that $t_{\rm
  pair}= -1.525$ may, in fact, be sufficient to weakly bind the two
{\em qps} if their clouds are fully converged and they are somewhat
slower/heavier. A definite answer will require significantly more
work, as more magnons will need to be added in the variational space
to fully converge the {\em qps'} clouds when they are far from each
other. We note that exact diagonalization of the same model on a
32Cu+64O cluster could not settle this issue either, because of
considerable finite-size effects, \cite{BayoB} although those results
also suggested that the system may be close to hosting pre-formed
pairs.

This issue clearly deserves further, careful study, which we plan to
attempt in the future. For now, we would like to speculate a bit
more on this topic, because what we do know is already quite interesting.

First, the fact that our parameters seem so close to the critical
region where pre-formed pairs may form suggests that on the BCS-BEC
spectrum, superconductivity mediated by this $H_{\rm{eff}}$ would be
more BEC than BCS-like, {\em i.e.} with pairs bound in real space, not
in momentum space. Of course, the possibility of cuprate
superconductivity emerging (at least on the underdoped side) when a
liquid of preformed pairs becomes coherent has long been one of the
leading scenarios.\cite{PF1,PF2,PF3,PF4,PF5,PF6}
More recently, several groups have suggested that various unusual
properties on the underdoped side can be explained as being due to the
scattering of fermionic carriers on a bosonic liquid of preformed
pairs.\cite{B1,B2,B3} Our
work seems to be consistent with these scenarios.

We leave it for future work to fully establish the symmetry of the
preformed pairs (if they exist) and/or of our effective attraction.
Note that the answer for the latter question is not trivial, because
of the many-band structure and because of the form of the effective
interaction. We can Fourier transform $H_{\rm eff}$, but (i) the
potential will depend not just on the momentum ${\bf q}$ exchanged
between the holes, but also on their total momentum ${\bf k+k'}$. More
importantly, (ii) because we have 4 different O sites in the magnetic
unit cell, this potential is in fact a $4\times 4$ matrix, and its
symmetry to rotations is more complicated to establish than for a
 scalar.

Even so, we can state that we expect this
symmetry to be $d$-wave like. The reason is as follows. We know that
for carriers moving on a square lattice like that made
by the O ions, any amount of on-site ($s$-wave symmetry) attraction
will lead to the appearance of a bound state. That we do not find this
bound state when $U_{pp} > 0$ may be explained by this  being larger than the $s$-wave component of the effective
attraction. However, we do not find a bound state even when $U_{pp}
=0$. This can only mean that our effective interaction does not have
a $s$-wave component, thus it is likely to be $d$-wave like.

\section{Summary and conclusions}

To conclude, we showed that magnon-exchange between holes doped in a
cuprate layer leads to an effective attraction, and identified its
expression $H_{\rm{eff}}$ and its energy scale $|t_{\rm pair}| \sim
1.5J_{dd}\sim 225$ meV.

The form of $H_{\rm{eff}}$ is unusual and requires further study. It
is interesting that it has a ``kinetic'' nature, as the holes move
while interacting. Evidence that pairing in cuprates comes through a
``kinetic'' mechanism was uncovered in optical experiments.\cite{O1-O5} As argued above, we also expect
$H_{\rm{eff}}$ to favor pairs with $d$-wave symmetry, and to be strong
enough so that the underdoped system either has pre-formed pairs, or
is very close to it. If pre-formed pairs exist, they are very weakly
bound, {\em i.e.} on a scale much smaller than $|t_{\rm pair}|$,
consistent with the fact that both $T_C$ and the pseudogap temperature
$T^*$ are well below $|t_{\rm pair}|$. Even if pre-formed pairs were
unstable, the superconductivity promoted by $H_{\rm{eff}}$ is likely
not BCS-like, but more towards BEC-like and unconventional.

As a final note, let us comment on why we expect this specific
magnon-mediated effective attraction $H_{\rm{eff}}$, derived in the
extremely underdoped limit and in the presence of LR AFM order, to be
relevant at least in the whole underdoped regime. The answer is that
this is interaction only involves the two holes and their common Cu
spin through which they exchange the magnon. To first order it makes
no difference whether this Cu spin is part of a magnetically ordered
system or not, especially as all energy scales characterizing
hole-spin interactions are much larger than $J_{dd}$. What may happen
with increased hole concentration is that the magnitude $|t_{\rm
  pair}|$ of this effective interaction is renormalized, but we expect
the functional form to remain the same.

Clearly, more work needs to be done to fully understand the
consequences of this specific $H_{\rm{eff}}$ attraction, but we
believe that the results reported here are interesting and intriguing, and warrant such further work.

\acknowledgments
We are grateful to G.~A. Sawatzky for many dicussions and suggestions. We acknowledge support from the Stewart Blusson Quantum Materials Institute and from the Natural Sciences and Engineering Research Council of Canada.

\appendix

\section{Hamiltonian details}
\label{app1}

\begin{figure}[b]
  \centering
  \includegraphics[width=0.4\columnwidth]{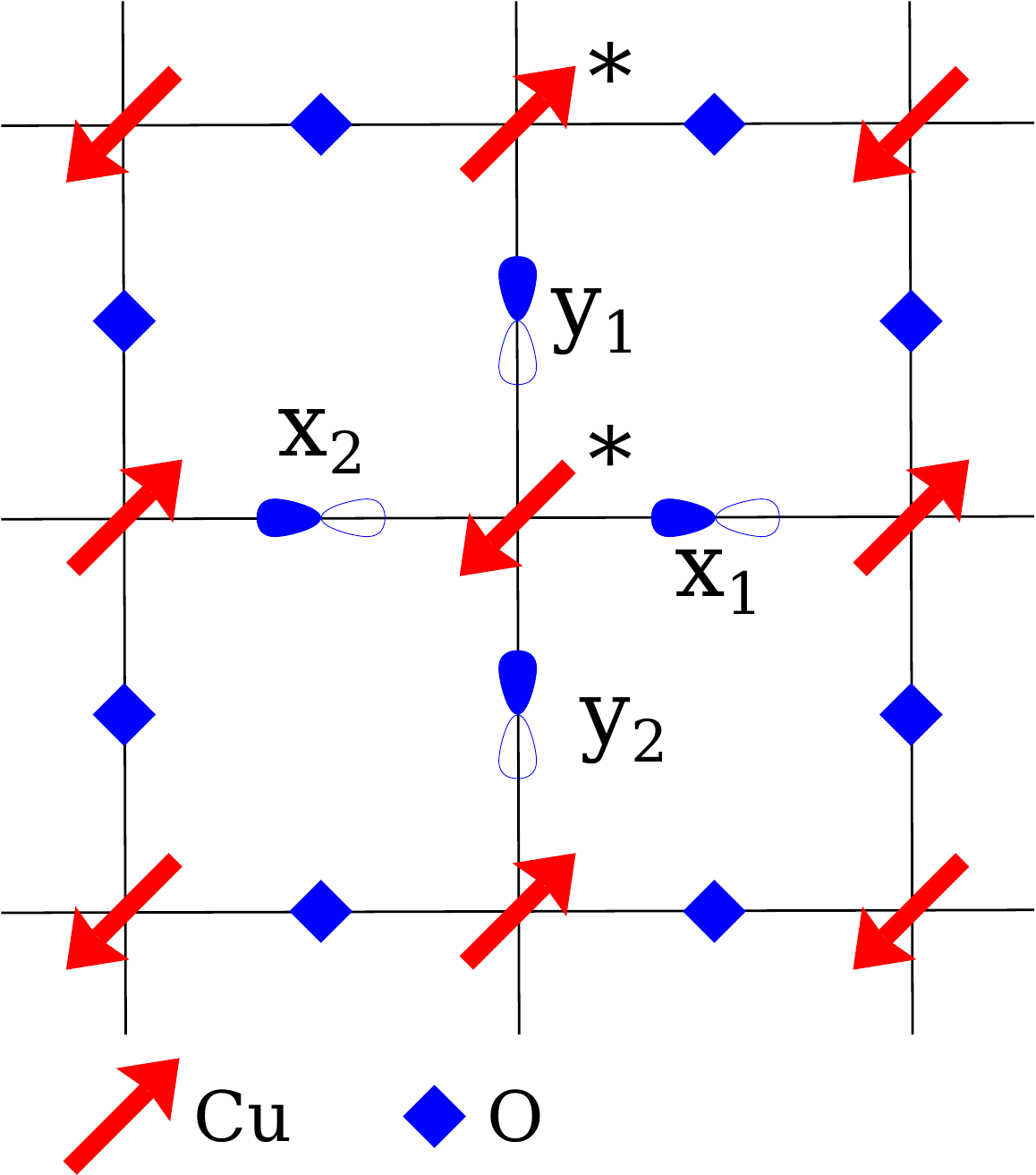}
  \caption{The CuO$_2$ lattice. The phase convention for the oxygen 2p orbitals
    is shown. Red arrows indicate the spin of the Cu-holes. The unit cell
    consists of the four oxygen sites marked with text and the two Cu-sites
    marked with an asterisk.}
  \label{fig:S1}
\end{figure}

We study the model introduced by Lau {\em et al.} \cite{Bayo} in the
two-hole sector, for a finite value of $U_{pp}$ (note that in Ref. \onlinecite{Bayo},
double occupancy is forbidden on the O sites). The holes propagate on the 2D
CuO$_2$-layer depicted in Fig. 1a of the main article. In the $U_{dd}
\rightarrow \infty$ limit at zero doping each Cu-ion is in a d$^9$
configuration, {\em i.e.} hosts a single hole which is described by a spin-degree of freedom. Due to
the superexchange interaction $ H_{J_{dd}}$ (see below), these Cu-spins tend to align
antiferromagnetically. As discussed in the main text, we first assume that this
interaction is of Ising type and thus, in the absence of doped holes,
the lattice of Cu-spins is in the N\' eel state. The
  role of the background spin-fluctuations enabled by the $x-y$ part of  $ H_{J_{dd}}$  is considered subsequently.

Starting from a  N\' eel state, the magnetic unit cell comprises two Cu-sites and four oxygen sites. Our choice
of unit cell is depicted in Fig. \ref{fig:S1}. For the $j^{\mr{th}}$ unit cell
the lattice vector $\mb{R}_j$ points to the Cu$_\dn$ site. The oxygen orbital
$\alpha$ is located at $\mb{r}_\alpha = \mb{R}_j + \mb{u}_{\alpha}$ and the
Cu$_\up$ site at $\mb{R}_j + a \hat{y}$, where $a$ is the lattice constant. It
is convenient to measure distances in units of $a$/2, as we do from here on.
Occasionally it will be convenient to sum over all the Cu$_\up$ sites. We will
indicate this by $j \in \mr{Cu}_\up$. In that case it is assumed that the vector
$\mb{R}_j$ points to a Cu$_\up$ site. It then follows (see Fig. \ref{fig:S1})
that the neighboring oxygen orbital of type $\alpha$ is located at $\mb{R}_j -
\mb{u}_\alpha$. If not otherwise stated, $\mb{R}_{j}$ is always assumed to
point to a Cu$_\dn$ site.

The two additional holes are
hosted by the ligand oxygen $2p$-orbitals pointing towards the nearest Cu-ions.
Their kinetic energy $T_{pp}$ is given by a tight binding Hamiltonian
describing nearest neighbor (NN) and next nearest neighbor (NNN) hopping across Cu sites:
 \begin{widetext}
\begin{equation}
\EqLabel{eq:Tpp}
T_{pp} =  t_{pp} \sum_{j,\sigma,\alpha,\mb{\delta}}
              \mr{s}(\mb{\delta}) c_{j+\mb{u}_{\alpha}+\mb{\delta},\sigma}^\dagger
              c_{j+\mb{u}_{\alpha},\sigma}  
            - t_{pp}^\prime \sum_{j,\sigma,\alpha} (c_{j+3\mb{u}_{\alpha},\sigma}^\dagger
              + c_{j-\mb{u}_{\alpha},\sigma}^\dagger) c_{j+\mb{u}_{\alpha},\sigma} + \mr{H.c.}
\end{equation}
 \end{widetext}
Here $c_{j+\mb{u}_{\alpha},\sigma}^\dagger$ ($ c_{j+\mb{u}_{\alpha},\sigma}$)
creates (annihilates) a hole with spin $\sigma$ at site $\mb{R}_{j} +
\mb{u}_{\alpha}$. The vectors $\mb{\delta}$ point to the four oxygen NN and
$\mr{s}(\mb{\delta})$ is the sign of the corresponding hopping amplitude, listed in
Tab. \ref{tab:hopping}. These signs are for holes (not electrons) and can be
inferred from the phases of the oxygen 2p orbitals depicted in Fig. \ref{fig:S1}.
The positive constants $t_{pp}$ and $t_{pp}^\prime$ are the magnitudes of the
NN and NNN hopping, respectively.

\begin{table}[b]
  \centering
  \begin{tabular}{|c|c|c|}
   $\mb{\delta}, \mb{\eta}$& $T_{pp}$ & $T_{sw}$  \\ \hline
    (1,1) & + & -\\
    (-1,1) & - & +\\
    (-1,-1) & + & -\\
    (1,-1) & - & +\\
    (2,0) & - & -\\
    (-2,0) & - & -\\
    (0,2) & - & -\\
    (0,-2) & - & -\\
  \end{tabular}
  \caption{Hopping signs for $T_{pp}$ and $T_{sw}$. The
    vectors $\mb{\delta}$ and $\mb{\eta}$ are given in units of $a$/2. The first
  four rows are the NN hopping directions, while the NNN directions are listed
  in the rows below.}
  \label{tab:hopping}
\end{table}

\begin{table}[b]
  \centering
  \begin{tabular}{|c|c|}
    orbital $\alpha$ & $\eta_\alpha$\\ \hline
    x1 & (-1,1) ; (-1,-1) ; (-2,0)\\
    x2 & (1,1) ; (1,-1) ; (2,0)\\
    y1 & (-1,-1) ; (1,-1) ; (0,-2)\\
    y2 & (-1,1) ; (1,1) ; (0,2)\\         
  \end{tabular}
  \caption{The vectors $\eta_\alpha$ pointing from orbital $\alpha$ to the
    oxygen orbitals which share a Cu$_\dn$ neighbor (in units of $a$/2).}
  \label{tab:Tswap_dirs}
\end{table}

The interaction between holes and Cu-spins has two terms.
The first is an exchange interaction:
\begin{align}
  \label{eq:HJpd}
  H_{J_{pd}} = \sum_{\alpha} \left( \sum_{j \in \mr{Cu}_\dn} + \sum_{j \in \mr{Cu}_\up}
  \right ) \vec{s}_{j+\mb{u}_{\alpha}} \cdot \vec{S}_j,
\end{align}
where $\vec{s}$ is the spin-operator for corresponding O holes and $\vec{S}$ is the
spin-operator for the Cu-spins.
The second term  involves
hopping of a hole while swapping its spin with the adjacent Cu:
\begin{widetext}
  \begin{align}
  \label{eq:Tsw}
    T_{\rm sw} &= t_{sw} \sum_{j \in \mr{Cu}-\dn} \sum_{\alpha,\sigma}
                            \mr{s}(\eta_{\alpha}) \left [  
                            c_{j+\mb{u}_\alpha + \eta_\alpha,-\sigma}^\dagger S_{j}^{\sigma}
                            (\frac{1}{2}- S_{j}^z \sigma)
                            + c_{j+\mb{u}_\alpha + \eta_\alpha,\sigma}^\dagger
                            (\frac{1}{2} + S_{j}^z \sigma) \right ] 
                            c_{j+\mb{u}_\alpha,\sigma} \nonumber \\
                         & +t_{sw} \sum_{j \in \mr{Cu}-\up} \sum_{\alpha,\sigma}
                            \mr{s}(\eta_{\alpha}) \left [  
                            c_{j-\mb{u}_\alpha - \eta_\alpha,-\sigma}^\dagger S_{j}^{\sigma}
                            (\frac{1}{2}- S_{j}^z \sigma)
                            + c_{j-\mb{u}_\alpha - \eta_\alpha,\sigma}^\dagger
                            (\frac{1}{2} + S_{j}^z \sigma) \right ] 
                            c_{j-\mb{u}_\alpha,\sigma}
\end{align}
\end{widetext}

Here $S_{j}^{\pm}$ are the ladder operators for the Cu-spins. The
vectors $\eta_\alpha$ point from orbital $\alpha$ to the other three oxygen
orbitals adjacent to the same Cu$_\dn$ site, see Tab. \ref{tab:Tswap_dirs}.
The vectors $-\eta_\alpha$ point to the three O which share a
 Cu$_\up$ site with the orbital $\alpha$. Note, furthermore, that for
$j \in \mr{Cu}_\up$, the vector $-\mb{u}_\alpha$ points to orbital $\alpha$.

When on-site Coulomb interaction $U_{pp}$ between the holes is included, the magnitude of 
$t_{sw}$ for the terms which involve a doubly-occupied site as either the
start or the final state, is renormalized by a factor $\frac{\Delta_{pd} +  U_{pp}/2}{\Delta_{pd} + U_{pp}}$, where $\Delta_{pd}$ is the charge transfer
gap. A derivation
of this renormalization can be obtained using the perturbation theory as in 
Ref. \onlinecite{Takahashi}.

In the absence of holes, the Cu-Cu Ising interaction
$\hat{H}_{J_{dd}}$ is given by
\begin{align}
  \label{eq:HJdd}
  H_{J_{dd}} = J_{dd} \sum_{j} \sum_{\alpha} S_{j}^z S_{j+2\mb{u}_\alpha}^z
\end{align}
In the presence of doped holes it vanishes for those Cu-pairs which
have one or more holes sitting on the O between them. In other words,
the holes block the magnetic superexchange.

 As mentioned, $ H_{J_{dd}}$ is in fact of Heisenberg,
  not Ising type. The difference is that the former promotes an
  undoped ground-state which contains background spin-fluctuations,
  while the later has a N\'eel ground-state without any background
  spin-fluctuations. In a later section, we will show that these spin
  fluctuations have no effect on the magnon-mediated effective
  interaction between holes. This is achieved by allowing
  spin-fluctuations to occur in the vicinity of holes (effectively
  restoring $ H_{J_{dd}}$ to its full Heisenberg form locally) and
  seeing if/how this affects the magnon-exchange. The first step, however, is
  to assume that there are no spin-fluctuations allowed, which we do from now on
  until specified otherwise.

Finally, including an on-site Hubbard interaction $U_{pp}$ between the
holes, we arrive at the total Hamiltonian:
\begin{align}
  \label{eq:Htot}
  \mathcal{H} = T_{pp} + U_{pp} + T_{\rm sw} + H_{J_{pd}} + H_{J_{dd}}
\end{align}
We use the same parameters as in Ref. \onlinecite{Bayo}, which in units of
$J_{dd}$ are $t_{pp} = 4.13$, $t_{pp}^\prime = 0.58 t_{pp}$, $t_{sw} = 2.98$
(2.20 for doubly occupied oxygen sites), $J_{pd} = 2.83$, $U_{pp} = 25.4$,
$\Delta_{pd} = 22.87$.

\section{Variational space details}
\label{app2}

To find the low-energy eigenstates we use a variational approach. This means
that we restrict the Hilbert space to a physically meaningful subspace, termed the
variational space (VS). In this variational space we can find the eigenstates, eigenenergies
and Green's functions using standard methods such as {\em e.g.}
the Lanczos algorithm.

The first restriction imposed on the VS is the maximum number of magnons
allowed, $n_m$. We are describing the Cu-spins as having an Ising exchange, so
the magnons are dispersionless and correspond to Cu-spins which are flipped with
respect to the N\'eel order (see main text discussion). For the single-hole case it was shown
\cite{HadiNP,HadiJCP} that reasonable convergence is already reached at $n_m =2$, because  every magnon costs a finite energy of order $J_{dd}$, yet  the magnons move very slowly compared to the
holes, allowing us to neglect their dispersion.

Considering only states with up to two magnons we define the following
translationally invariant basis states which span the VS:
\begin{align}
  &|0,k, \mb{R}, \mb{u}_{\alpha} , \mb{u}_{\beta} \rangle = \sum_{j} \frac{\e{i \mb{k} \mb{R}_j}}{\sqrt{N}}
    c_{j+\mb{u}_\alpha,\up}^\dagger c_{j+\mb{R}+\mb{u}_\beta,\dn}^\dagger |0\rangle
  \nonumber \\
  &|\sigma, k, \mb{r}_{\alpha} , \mb{r}_{\beta} \rangle = \sum_{j}
    \frac{\e{i \mb{k} \mb{R}_j}}{\sqrt{N}}
    c_{j+\mb{r}_\alpha,-\sigma}^\dagger c_{j+\mb{r}_\beta,-\sigma}^\dagger
    S_{j}^\sigma|0\rangle
  \nonumber \\
  \label{eq:basis-states}
  &|2, k, \mb{R}, \mb{r}_{\alpha} , \mb{r}_{\beta} \rangle = \sum_{j}
    \frac{\e{i \mb{k} \mb{R}_j}}{\sqrt{N}}
    c_{j+\mb{r}_\alpha,\up}^\dagger c_{j+\mb{r}_\beta,\dn}^\dagger
    S_{j}^+S_{j+\mb{R}+2\hat{y}}^- |0\rangle
\end{align}
Here $|0\rangle$ is the undoped N\'eel state and $N \rightarrow \infty$ denotes the number of lattice sites. All other quantities were defined in the preceding
Section.

For all these states the distance between holes and/or magnons is
well defined. For the zero-magnon and two-magnon states the holes have opposite
spin and are therefore distinguishable. This is not true for the one-magnon
states. In order to not double-count one-magnon states we require that
$\mb{r}_\alpha$ is lexicographically smaller than $\mb{r}_\beta$.

For the zero-magnon states the reference unit cell is that of the $\up$-hole,
for the $\up$-magnon and for the two-magnon states it is that of the $\up$-magnon, and
for the $\dn$-magnon states it is that of the $\dn$-magnon. These choices are
convenient because  the magnons do not move when we ignore the background spin-fluctuations.

To get a numerical solution, we need to further restrict the size of the VS. This is
achieved by introducing two more cutoffs which are calculated using the L1 norm. The
first is denoted by $M_c$ and restricts the distance between any two
particles (particle refers to both holes and magnons). The second cutoff
$m_c \leq M_c$ restricts the distance between a magnon and its
closest hole. For example, for the zero-magnon states we require that
$||\mb{R}+\mb{u}_\beta - \mb{u}_\alpha||_{1} \leq M_c$. Because no magnons are
present, $m_c$ is irrelevant for the zero-magnon states.

A slightly more complicated example are the $\up$-magnon states. The restriction
$||\mb{R_2}+\mb{u}_\beta - \mb{R_1} -\mb{u}_\alpha||_{1} \leq M_c$ is always
enforced. Furthermore one of the following sets of restrictions must also be enforced
{\em (i)} $||\mb{R_1} + \mb{u}_\alpha||_{1} \leq m_c$ and $||\mb{R_2} +
\mb{u}_\beta||_{1} \leq M_c$, or {\em (ii)} $||\mb{R_2} + \mb{u}_\beta||_{1} \leq
m_c$ and $||\mb{R_1} + \mb{u}_\alpha||_{1} \leq M_c$.
For the two-magnon states the restrictions are imposed in the same manner. Note
that in this case it is possible that both magnons are within 
$m_c$ of the same hole.

As discussed in the main text, we turn off the magnon-mediated
interactions by labelling the holes and magnons as being either flavor
$a$ or $b$, and allowing each flavor of hole to interact only with its
own flavor of magnon. Physical restrictions such as not allowing two
magnons at the same Cu site are imposed, as further discussed below. The
resulting states can easily be generalized from those in Eq.
\eqref{eq:basis-states}, as shown in Fig 1. (d) and (e).

\begin{figure}
  \centering
  \includegraphics[width=\columnwidth]{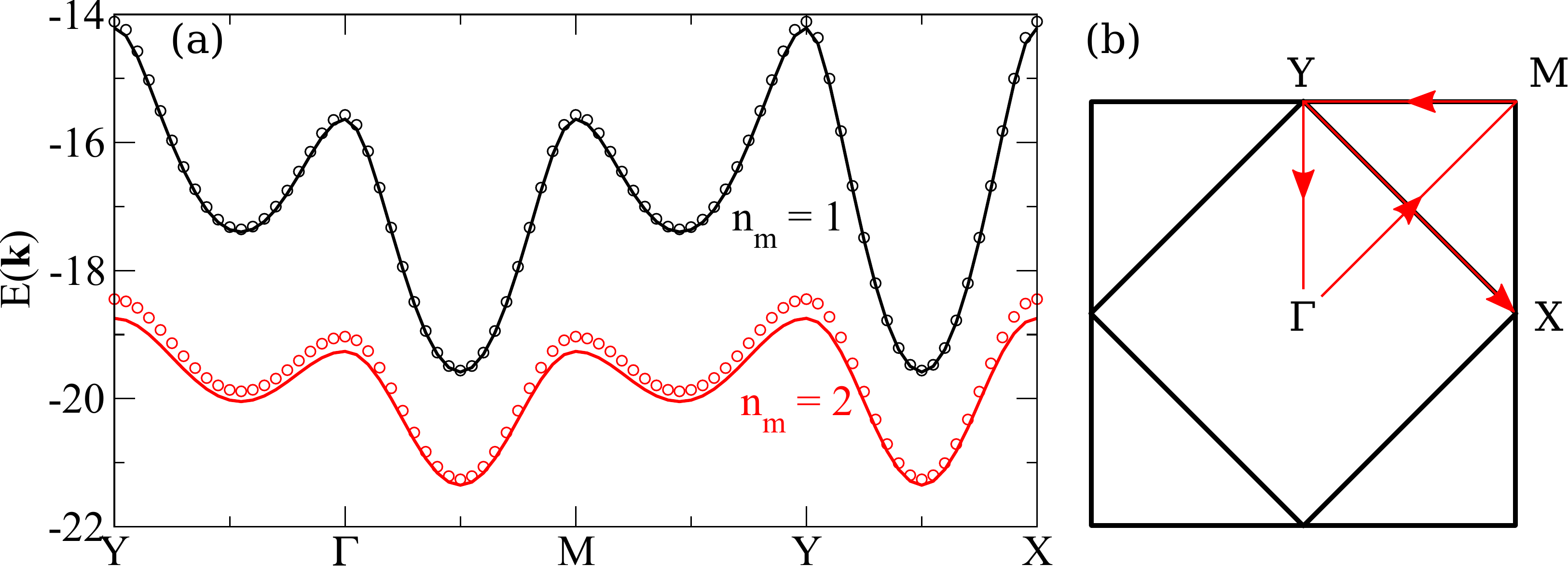}
  \caption{(a) The dispersion of the single hole along high symmetry lines of
    the BZ for a cutoff of $n_m = 1 $ and $n_m = 2$. Lines are the fully
    converged solution ($m_c =20$) and circles are results for $m_c = 3$. (b) A
    sketch of the cut in the BZ. The magnetic BZ is the small square.}
  \label{fig:S2}
\end{figure}

For completeness,  we now quickly review some aspects of the
single-hole solution. The single-hole results shown here are identical to those
from Refs \onlinecite{HadiNP,HadiJCP}. The low-energy quasiparticle
(QP) is a spin-polaron, {\it i.e.} a state in which the hole coherently emits
and reabsorbs magnons. In Fig. \ref{fig:S2} we show the {\em qp} dispersion,
$\epsilon_{\mr{sp}}(\mb{k})$, along high symmetry lines of the BZ. Note that due
to the AFM order of the Cu-spins, the BZ is reduced as shown in panel (b)
of Fig. \ref{fig:S2}. As a result the $\Gamma$ and $M$ points are equivalent, as are  X and Y.

For the single-hole solution we only need two cutoffs: $m_c$ (the hole-magnon
distance) and $n_m$ (the maximum number of magnons). The solid lines in Fig.
\ref{fig:S2} are fully converged in $m_c$ while the open circles are for $m_c =
3$. The effect of increasing $n_m$ from 1 to 2 is a constant energy shift and a
decrease in bandwidth, while the shape of the dispersion remains similar. The
single-hole calculation is essentially converged at $n_m = 2$.
\cite{HadiNP,HadiJCP}

%

\section{Derivation of $H_{\mr{eff}}$}  
\label{app3}

As described in the main text, we use second order perturbation theory (PT) to provide  guidance for the possible types of terms that may arise when a magnon is exchanged between the two holes. To make sure that the magnon is truly exchanged, we have to work in the enlarged variational space where the holes and magnons have flavors so that we can distinguish direct processes (whereby the same hole creates and absorbs the magnon) from the exchange ones (where the magnon is created by one hole and absorbed by the other). We then project back to the physical space with $c, c^\dagger$ operators by using the physical antisymmetrical combinations that enforce Pauli's principle. For example, the zero-magnon states are related by (also see Eq. \ref{eq:basis-states}):
\begin{widetext}
\begin{align}
  |0,\mb{k},\mb{R},\mb{u}_\alpha,\mb{u}_\beta \rangle &\leftrightarrow \sum_{j}
  \frac{\e{i \mb{k} \mb{R}_j}}{\sqrt{2N}}
  (a^\dagger_{j+\mb{u},_\alpha\uparrow} b^\dagger_{j + \mb{R}+\mb{u}_\beta, \downarrow} - a^\dagger_{j + \mb{R}+\mb{u}_\beta, \downarrow} b^\dagger_{j+\mb{u},_\alpha\uparrow} )
  | 0 \rangle 
  \nonumber \\
  &\equiv \frac{1}{\sqrt{2}} [|0,\mb{k},\mb{R},\mb{u}_\alpha \up ,\mb{u}_\beta \dn\rangle_E
  - \e{- i \mb{k} \mb{R}}
    |0,\mb{k},-\mb{R},\mb{u}_\beta \dn ,\mb{u}_\alpha \up \rangle_E ]
    \label{eq:c-ab-conversion}
\end{align}
where  for convenience,   we defined the zero-magnon states in the extended variational space:
\begin{equation}
  |0,\mb{k},\mb{R},\mb{u}_\alpha \sigma, \mb{u}_\beta -\sigma \rangle_E  =
  \sum_{j} \frac{\e{i \mb{k} \mb{R}_j}}{\sqrt{N}}
  a^\dagger_{j+ \mb{u}_\alpha, \sigma} b^\dagger_{j+ \mb{R}+\mb{u}_\beta, -\sigma}
  | 0 \rangle
\end{equation}

\end{widetext}

\begin{figure}[b]
  \centering
  \includegraphics[width=0.9\columnwidth]{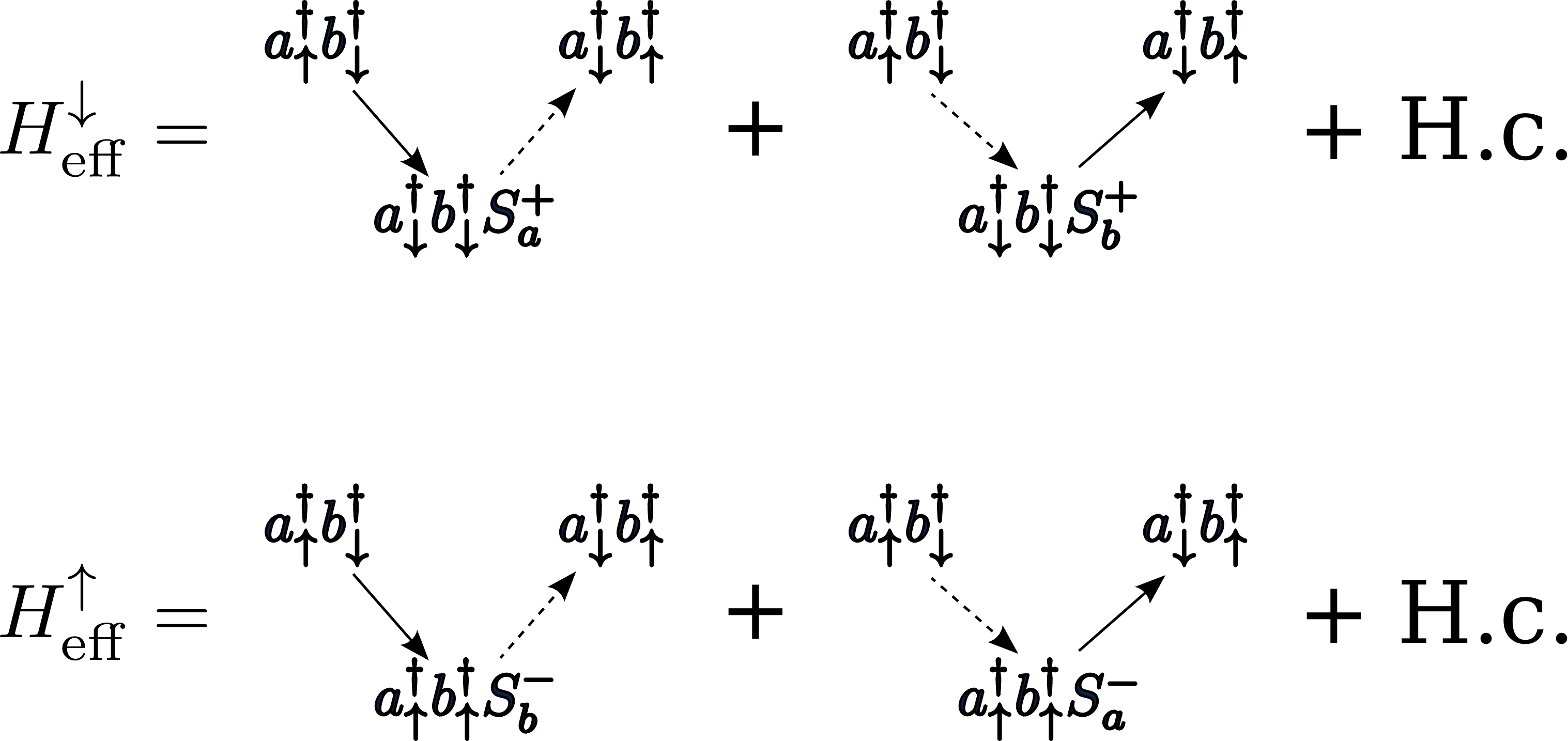}
  \caption{Sketch of the terms included in $H_{\mr{eff}}$. Solid (dashed)
    lines indicate normal (exchange) emission or absorption.}
  \label{figS6}
\end{figure}

Next, we note that $H_{\mr{eff}}$ must be of the form $H_{\mr{eff}}
= H_{\mr{eff}}^{\up} + H_{\mr{eff}}^{\dn}$, where the superscript
indicates whether an (originally)  Cu$_\up$ or Cu$_\dn$ is mediating the magnon
exchange. These terms are schematically depicted in Fig. \ref{figS6}.
Here we only show the terms which take a state from the $a_{\up}
b_{\dn}$-family to the $a_{\dn}b_{\up}$-family. The other possible terms are
just their Hermitian conjugates.

Finally, as mentioned in the main text, there are four possible kinds
of terms in the full $H_{\mr{eff}}$, depending on whether the magnon
is emitted and absorbed through $T_{\rm sw}$ and/or $J_{pd}$
processes (where one process is direct and the other is exchange). We
have derived all these terms and analyzed each individually, as
discussed in the main text. Because the term where both processes are
of $T_{\rm sw}$ type turns out to dominate and to provide a faithful
description of the magnon-exchange effects in the full Brillouin zone,
in the following we focus on its derivation. All the other terms can
be derived similarly.

\subsection{Derivation of $H_{\mr{eff}}^{\dn}$}

Due to our choice of unit cell, it is easier to deal with
$H_{\mr{eff}}^\dn$. The two terms depicted in Fig. \ref{figS6} only
differ by the magnon-label of the intermediate state. They therefore
give the same result and we only need to consider the first term,
which corresponds to first emitting with $T_{\mr{sw}}^{\mr{d}}$ and
then reabsorbing with $T_{\mr{sw}}^{\mr{e}}$, where $d/e$ labels are
for direct/exchange processes. Consequently, the effect of
$H_{\mr{eff}}^\dn$ on the state $|0,\mb{k},\mb{R},\mb{u}_\alpha \up,
\mb{u}_\beta \dn \rangle_E$ is (see Eq. \eqref{eq:Tsw}):
\begin{widetext}
\begin{align}
  H_{\mr{eff}}^{\dn} |0,\mb{k},\mb{R},\mb{u}_\alpha \up, \mb{u}_\beta \dn \rangle_E =
  2 t_{\mr{pair}} \delta_{\mb{R},0} \sum_{\substack{\eta_\alpha \neq \\ \mb{u}_\beta -
      \mb{u}_\alpha}} \sum_{\eta_\beta} \mr{s}(\eta_{\alpha}) \mr{s}(\eta_\beta)
  |0,\mb{k},\mb{R},\mb{u}_\alpha + \eta_{\alpha} \dn,\mb{u}_\beta + \eta_{\beta} \up \rangle_E
\end{align}
Here, as discussed in the main text, we leave $t_{\mr{pair}}$ as a parameter to be fitted, instead of using its PT expression. 
The restriction
on $\eta_\alpha$ ensures that in the intermediate state where the holes have
the same spin, they are not on the same site. Furthermore the holes
need to sit on the ``cage'' surrounding a Cu-$\dn$ ion, which leads to
the appearance of $\delta_{\mb{R},0}$. The result for the state
$|0,\mb{k},-\mb{R},\mb{u}_\beta \dn ,\mb{u}_\alpha \up \rangle_E$ is derived similarly. Using the relationship of Eq. \ref{eq:c-ab-conversion} between states in the extended variational space and those in the physical space, we then find: 
\begin{align}
  & H_{\mr{eff}}^\dn  \sum_{j} \frac{\e{i \mb{k}\mb{R}_j}}{\sqrt{N}} c_{j+\mb{u}_\alpha,\up}^\dagger
                     c_{j+\mb{R}+\mb{u}_\beta,\dn}^\dagger |0 \rangle
   \nonumber \\
  &  = 2 t_{\mr{pair}} \delta_{\mb{R},0} 
  \sum_{\substack{\eta_\alpha \neq \\ \mb{u}_\beta - \mb{u}_\alpha}}
  \sum_{\eta_\beta} \mr{s}(\eta_{\alpha}) \mr{s}(\eta_\beta)\frac{1}{\sqrt{2}} [ 
  |0,\mb{k},\mb{R},\mb{u}_\alpha + \eta_{\alpha} \dn,\mb{u}_\beta + \eta_{\beta} \up \rangle_E
  - \e{ -i \mb{k}\mb{R}}
  |0,\mb{k},\mb{-R},\mb{u}_\beta + \eta_{\beta} \up,\mb{u}_\alpha + \eta_{\alpha} \dn \rangle_E
    ] \nonumber \\
 &  = - 2 t_{\mr{pair}} \delta_{\mb{R},0} 
  \sum_{\substack{\eta_\alpha \neq \\ \mb{u}_\beta - \mb{u}_\alpha}}
  \sum_{\eta_\beta} \mr{s}(\eta_{\alpha}) \mr{s}(\eta_\beta) 
 \frac{\e{ - i \mb{k} \mb{R}}}{\sqrt{2}} [
  |0,\mb{k},\mb{-R},\mb{u}_\beta + \eta_{\beta} \up,\mb{u}_\alpha + \eta_{\alpha} \dn \rangle_E                     
  - \e{i \mb{k}\mb{R}}
  |0,\mb{k},\mb{R},\mb{u}_\alpha + \eta_{\alpha} \dn,\mb{u}_\beta + \eta_{\beta} \up \rangle_E
    ] \nonumber \\
&  =  - 2 t_{\mr{pair}} \delta_{\mb{R},0} 
  \sum_{\substack{\eta_\alpha \neq \\ \mb{u}_\beta - \mb{u}_\alpha}}
  \sum_{\eta_\beta} \mr{s}(\eta_{\alpha}) \mr{s}(\eta_\beta) \e{ -i \mb{k} \mb{R}}
  \sum_j \frac{\e{i \mb{k}\mb{R}_j}}{\sqrt{N}} c_{j+\mb{u}_\beta + \eta_\beta,\up}^\dagger
  c_{j-\mb{R}+\mb{u}_\alpha + \eta_\alpha,\dn}^\dagger |0 \rangle
\end{align}
This holds for any choice of $\alpha$ and $\beta$, therefore we can immediately
read off that $H_{\mr{eff}}^\dn$ must have the form listed in the main text:
\begin{align}
  H_{\mr{eff}}^\dn = -2 t_{\mr{pair}} \delta_{\mb{R},0} \sum_j \sum_{\alpha,\beta}
  \sum_{\substack{\eta_\alpha \neq \\ \mb{u}_\beta - \mb{u}_\alpha}}
  \sum_{\eta_\beta} \mr{s}(\eta_{\alpha}) \mr{s}(\eta_\beta)
  c_{j+\mb{u}_\beta + \eta_\beta, \up}^\dagger c_{j+\mb{u}_\alpha + \eta_\alpha, \dn}^\dagger
  c_{j+\mb{u}_\beta, \dn} c_{j+\mb{u}_\alpha, \up} 
\end{align}

\subsection{Derivation of $H_{\mr{eff}}^{\up}$}

To obtain an expression for $H_{\mr{eff}}^{\up}$ we first rewrite 
$|0,\mb{k},\mb{R},\mb{u}_\alpha \up, \mb{u}_\beta \dn \rangle_E\rangle$ (where the sum is over Cu$_\dn$ sites by definition) 
so that the sum is
over Cu$_\up$ sites instead. The Cu$_\up$ site at $\mb{R}_j+2\mb{u}_\alpha$ is closest
to the 'a' hole at $\mb{R}_j+\mb{u}_\alpha$. Consequently we make the
substitution $R_{j'} = \mb{R}_j + 2\mb{u}_\alpha$ which yields
\begin{align}
  |0,\mb{k},\mb{R},\mb{u}_\alpha \up, \mb{u}_\beta \dn \rangle_E  = \sum_{j' \in \mr{Cu}_\up}
\frac{\e{i \mb{k}( \mb{R}_{j'} - 2\mb{u}_\alpha)}}{\sqrt{N}}
\phi_{a,\up}(\mb{R}_{j'}-\mb{u}_\alpha)
\phi_{b,\dn}(\mb{R}_{j'}+\mb{R}-2\mb{u}_\alpha+2\mb{u}_\beta - \mb{u}_\beta) | 0 \rangle.
\end{align}
\end{widetext}

Before we continue the following observations are helpful. The site $\mb{R}_{j'}
- \mb{u}_\alpha$ is still the site of an $\alpha $ orbital. Furthermore the
vector $2 \mb{u}_{\alpha} - 2 \mb{u}_{\beta}$ is a lattice vector so that we
must have $\mb{R} = 2 \mb{u}_{\alpha} - 2 \mb{u}_{\beta} $ in order for the two
holes to share the same Cu$_\up$ neighbor.

To calculate the effect of $H_{\mr{eff}}^\up$ we make use of the fact that from
an orbital of type $\alpha$, the vectors $-\eta_{\alpha}$ point to the sites
which can be reached by ``hopping over'' the NN Cu$_\up$ site. We then obtain:
\begin{widetext}
\begin{align}
&  H_{\mr{eff}}^{\up} |0,\mb{k},\mb{R},\mb{u}_\alpha \up, \mb{u}_\beta \dn \rangle_E 
   \nonumber \\ & 
= 
  2 t_{\mr{pair}} \delta_{\mb{R},2\mb{u}_\alpha - 2\mb{u}_\beta}
  \sum_{\substack{\eta_\beta \neq \\ \mb{u}_\alpha - \mb{u}_\beta}} \sum_{\eta_{\alpha}}
  \mr{s}(-\eta_\beta) \mr{s}(-\eta_\alpha) \sum_{j' \in \mr{Cu}_\up}  \frac{\e{i \mb{k}( \mb{R}_{j'} - 2\mb{u}_\alpha)}}{\sqrt{N}}
  \phi_{a,\dn}(\mb{R}_{j'}-\mb{u}_\alpha - \eta_\alpha)
  \phi_{b,\up}(\mb{R}_{j'} - \mb{u}_\beta - \eta_{\beta}) | 0 \rangle.
\end{align}
We now have to rewrite this back as a sum over Cu$_\dn$ sites. To do this we
make the substitution $\mb{R}_j = \mb{R}_{j'} - 2 \mb{u}_\alpha - 2
\eta_\alpha$. This gives:
\begin{align}
&  H_{\mr{eff}}^{\up} |0,\mb{k},\mb{R},\mb{u}_\alpha \up, \mb{u}_\beta \dn \rangle_E   \nonumber \\
& =
  2 t_{\mr{pair}} \delta_{\mb{R},2\mb{u}_\alpha - 2\mb{u}_\beta}
  \sum_{\substack{\eta_\beta \neq \\ \mb{u}_\alpha - \mb{u}_\beta}} \sum_{\eta_{\alpha}}
  e^{2 i \mb{k} \eta_\alpha} \mr{s}(-\eta_\beta) \mr{s}(-\eta_\alpha)   | 0, \mb{k},\mb{R} + 2 \eta_\alpha - 2 \eta_\beta, \mb{u}_\alpha + \eta_\alpha \dn,
  \mb{u}_\beta + \eta_\beta, \up \rangle_E
\end{align}

A similar calculation yields
\begin{align}
&  H_{\mr{eff}}^\up |0, \mb{k},-\mb{R}, \mb{u}_\beta\dn, \mb{u}_\alpha \up \rangle_E
 \nonumber \\
    &  = 2 t_{\mr{pair}} \delta_{\mb{R},2\mb{u}_\alpha-2 \mb{u}_\beta}
  \sum_{\substack{\eta_\beta \neq \\ \mb{u}_\alpha - \mb{u}_\beta}} \sum_{\eta_{\alpha}}
  \mr{s}(-\eta_\beta) \mr{s}(-\eta_\alpha) \e{2 i \mb{k} \eta_\beta} |0,\mb{k}, -\mb{R} +2 \eta_\beta - 2\eta_\alpha,
      \mb{u}_\beta + \eta_\beta \up, \mb{u}_\alpha + \eta_\alpha \dn \rangle_E
\end{align}

Making use of Eq. \eqref{eq:c-ab-conversion}, the effect of $H_{\mr{eff}}^\up$ in
the language of the $c$-operators is:
\begin{align}
  H_{\mr{eff}}^\up & \sum_j \e{i \mb{k} \mb{R}_j} c_{j+\mb{u}_\alpha, \up}^\dagger
  c_{j+\mb{R}+\mb{u}_\beta, \dn}^\dagger |0\rangle \nonumber \\
  = &2 t_{\mr{pair}} \delta_{\mb{R},2\mb{u}_\alpha-2 \mb{u}_\beta}
  \sum_{\substack{\eta_\beta \neq \\ \mb{u}_\alpha - \mb{u}_\beta}} \sum_{\eta_{\alpha}}
  \mr{s}(-\eta_\beta) \mr{s}(-\eta_\alpha) 
  \frac{1}{\sqrt{2}} [ \e{2 i \mb{k} \eta_\alpha}|0, \mb{k}, \mb{R} + 2\eta_\alpha - 2 \eta_\beta,
  \mb{u}_\alpha + \eta_\alpha \dn, \mb{u}_\beta + \eta_\beta \up \rangle_E \nonumber \\
  & - \e{2 i \mb{k} \eta_\beta-i \mb{k} \mb{R}} |0,\mb{k}, -\mb{R} +2 \eta_\beta - 2\eta_\alpha,
    \mb{u}_\beta + \eta_\beta \up, \mb{u}_\alpha + \eta_\alpha \dn \rangle_E ] \nonumber \\
  = &- 2 t_{\mr{pair}} \delta_{\mb{R},2\mb{u}_\alpha-2 \mb{u}_\beta}
  \sum_{\substack{\eta_\beta \neq \\ \mb{u}_\alpha - \mb{u}_\beta}} \sum_{\eta_{\alpha}}
  \mr{s}(-\eta_\beta) \mr{s}(-\eta_\alpha)
  \frac{\e{2 i \mb{k} \eta_\beta -i\mb{k} \mb{R} }}{\sqrt{2}}
  [|0,\mb{k}, -\mb{R} +2 \eta_\beta - 2\eta_\alpha,
  \mb{u}_\beta + \eta_\beta \up, \mb{u}_\alpha + \eta_\alpha \dn \rangle_E \nonumber \\
  & - \e{i \mb{k}( \mb{R} + 2\eta_\alpha - 2\eta_\beta)}
  |0, \mb{k}, \mb{R} + 2\eta_\alpha - 2 \eta_\beta, \mb{u}_\alpha + \eta_\alpha \dn,
    \mb{u}_\beta + \eta_\beta \up \rangle_E ] \nonumber \\
  = & - 2 t_{\mr{pair}} \delta_{\mb{R},2\mb{u}_\alpha-2 \mb{u}_\beta}
  \sum_{\substack{\eta_\beta \neq \\ \mb{u}_\alpha - \mb{u}_\beta}} \sum_{\eta_{\alpha}}
  \mr{s}(-\eta_\beta) \mr{s}(-\eta_\alpha) \e{2 i \mb{k} \eta_\beta -i\mb{k} \mb{R} }
  \sum_j \e{i \mb{k} \mb{R}_j} 
 c_{j+\mb{u}_\beta + \eta_\beta, \up}^\dagger
  c_{j-\mb{R} + 2 \eta_\beta - 2\eta_\alpha +\mb{u}_\alpha + \eta_\alpha, \dn}^\dagger |0\rangle
\end{align}

To read off an expression for $H_{\mr{eff}}^\up$ we transform the sums over $j$
on both sides of the equation so that $\mb{R}_{j} \in $ Cu$_\up$. For the sum on
the left hand side this is achieved with the substitution $\mb{R}_{j'} =
\mb{R}_{j} + 2 \mb{u}_\alpha$. For the sum on the right hand side we substitute
$\mb{R}_{j'} = \mb{R}_{j} + 2 \mb{u}_\beta + 2 \eta_\beta$.
\begin{align}
  &\sum_{j' \in \mr{Cu}_\up} H_{\mr{eff}}^\up \e{i \mb{k} (\mb{R}_{j'} - 2 \mb{u}_\alpha)}
  c_{j'-\mb{u}_\alpha, \up}^\dagger
  c_{j' +\mb{R} - 2\mb{u}_\alpha + 2\mb{u}_\beta - \mb{u}_\beta , \dn}^\dagger |0\rangle
    \nonumber \\
  =& - 2 t_{\mr{pair}} \delta_{\mb{R},2\mb{u}_\alpha-2 \mb{u}_\beta}
  \sum_{\substack{\eta_\beta \neq \\ \mb{u}_\alpha - \mb{u}_\beta}} \sum_{\eta_{\alpha}}
  \mr{s}(-\eta_\beta) \mr{s}(-\eta_\alpha) \e{2 i \mb{k} \eta_\beta -i\mb{k} \mb{R} }
  \sum_{j' \in \mr{Cu}_\up} \e{i \mb{k} ( \mb{R}_{j'} - 2 \mb{u}_\beta - 2 \eta_\beta)}
 c_{j'-\mb{u}_\beta - \eta_\beta, \up}^\dagger
  c_{j' - \mb{R} + 2\mb{u}_\alpha - 2 \mb{u}_\beta - \mb{u}_\alpha - \eta_\alpha, \dn}^\dagger |0\rangle
\end{align}
Note that when we make use of the $\delta$-function, $\mb{R}$ cancels out and the
terms in the exponential cancel as well, so that we obtain:
\begin{align}
  \sum_{j' \in \mr{Cu}_\up} \e{i \mb{k} \mb{R}_{j'}} [ H_{\mr{eff}}^\up 
  c_{j'-\mb{u}_\alpha, \up}^\dagger
  c_{j'  - \mb{u}_\beta , \dn}^\dagger |0\rangle
  +  2 t_{\mr{pair}} 
  \sum_{\substack{\eta_\beta \neq \\ \mb{u}_\alpha - \mb{u}_\beta}} \sum_{\eta_{\alpha}}
  \mr{s}(-\eta_\beta) \mr{s}(-\eta_\alpha) c_{j'-\mb{u}_\beta - \eta_\beta, \up}^\dagger
  c_{j'  - \mb{u}_\alpha - \eta_\alpha, \dn}^\dagger |0\rangle ] = 0
\end{align}
Consequently we must have
\begin{align}
  H_{\mr{eff}}^\up  = 2 t_{\mr{pair}} \sum_{j \in \mr{Cu}_\up} \sum_{\alpha,\beta}
  \sum_{\substack{\eta_\beta \neq \\ \mb{u}_\alpha - \mb{u}_\beta}} \sum_{\eta_{\alpha}}
  \mr{s}(-\eta_\beta) \mr{s}(-\eta_\alpha)
  c_{j-\mb{u}_\beta - \eta_\beta, \up}^\dagger
  c_{j  - \mb{u}_\alpha - \eta_\alpha, \dn}^\dagger
  c_{j  - \mb{u}_\beta , \dn}
  c_{j-\mb{u}_\alpha, \up}
\end{align}
Note that this expression is essentially the same as for $H_{\mr{eff}}^\dn$, but
with the first sum running over all the Cu$_\up$ sites instead of the Cu$_\dn$ sites. That is in agreement with what is expected by symmetry, validating these derivations.

\end{widetext}


\end{document}